\documentclass[sigconf]{acmart}

\usepackage[english]{babel}
\usepackage{blindtext}

\usepackage{graphicx}
\usepackage{amsmath}
\usepackage{xcolor}
\usepackage{mathtools}
\usepackage{amsfonts}
\usepackage{hyperref}
\usepackage{amsthm}

\usepackage{booktabs}
\usepackage{algorithmic}
\usepackage{algorithm}

\allowdisplaybreaks
\renewcommand\footnotetextcopyrightpermission[1]{} 
\setcopyright{none}

\acmDOI{}

\acmISBN{}

\acmPrice{}

\usepackage{graphicx}
\usepackage{amsmath}
\usepackage{xcolor}
\usepackage{mathtools}
\usepackage{amsfonts}
\usepackage{hyperref}
\usepackage{amsthm}

\usepackage{booktabs}
\usepackage{algorithmic}
\usepackage{algorithm}

\allowdisplaybreaks
\begin{document}

\title{Sparse Mean Field Load Balancing in Large Localized Queueing Systems}


\author{Anam Tahir, Kai Cui, Heinz Koeppl}

\renewcommand{\shortauthors}{X.et al.}

\begin{abstract}
    Scalable load balancing algorithms are of great interest in cloud networks and data centers, necessitating the use of tractable techniques to compute optimal load balancing policies for good performance. However, most existing scalable techniques, especially asymptotically scaling methods based on mean field theory, have not been able to model large queueing networks with strong locality. Meanwhile, general multi-agent reinforcement learning techniques can be hard to scale and usually lack a theoretical foundation. In this work, we address this challenge by leveraging recent advances in sparse mean field theory to learn a near-optimal load balancing policy in sparsely connected queueing networks in a tractable manner, which may be preferable to global approaches in terms of wireless communication overhead. Importantly, we obtain a general load balancing framework for a large class of sparse bounded-degree wireless topologies. By formulating a novel mean field control problem in the context of graphs with bounded degree, we reduce the otherwise difficult multi-agent problem to a single-agent problem. Theoretically, the approach is justified by approximation guarantees. Empirically, the proposed methodology performs well on several realistic and scalable wireless network topologies as compared to a number of well-known load balancing heuristics and existing scalable multi-agent reinforcement learning methods.
    
\end{abstract}

\maketitle

\section{Introduction}
The increasing demand for computational resources has led to increased research interest in parallel and distributed systems such as data centres and large-scale cloud networks, which can be accurately modelled by large-scale queueing networks due to their stochastic nature \cite{rodriguez2018cloud,  walker2022performance}. 
As a result, there is also a renewed interest in studying and designing scalable load balancing algorithms to allow these parallel systems to operate efficiently by reducing queue lengths, minimizing job waiting times, increasing system throughput, etc. \cite{mishra2020load}.
This work also focuses on load balancing policies in large networks with the goal of reducing overall job drops in the system.
Here, we will consider sparsely connected queueing systems,
rather than assuming that all agents can access to all the queues.
Many successful centralized and decentralized algorithms have already been proposed where the load balancer (agent) sends jobs to the available parallel $M$ servers (queues). These include (i) join-the-shortest-queue (JSQ) to reduce expected job delay when servers are homogeneous, have infinite buffer space, and independent, identical (i.i.d.), and exponentially distributed job service times \cite{winston1977optimality}, (ii) shortest-expected-delay (SED) to minimize the mean response time of jobs when service times are exponential, but with different service rates \cite{selen2016steady}, (iii) join-idle-queue (JIQ) to balance the overall system load while reducing communication overhead between load balancers and servers \cite{lu2011join}, and more, see e.g. \cite{der2022scalable, mukherjee2018asymptotically}.
However, the first two aforementioned algorithms are asynchronous and assume instantaneous knowledge of queue states, which is not true in practice, especially in large systems. And for JIQ, queues also need to be maintained at the scheduler end which we do not consider in our system model.


Often, large queueing systems
are modelled as a decentralized, multi-agent system with an underlying graph topology. 
Different types of graphs can be used to represent different types of (wireless) network topologies and information structures, for example a cube-connected cycle was  used to represent a versatile network of agents in parallel computing \cite{habibian2017fault}. 
The vertices in the graph represent agents (load balancers) and the edges represent the (possibly wirelessly connected) neighbourhood of each agent, resulting in local states, actions and information exchange.
In this work, all agents with access to the same queues form a neighbourhood, and each agent is assumed to only allocate incoming load to queues within their own neighbourhood.

\paragraph*{Related work} 
The aforementioned decentralized queueing systems with underlying graph structure can be modelled as various variants of multi-agent (partially-observable) Markov decision processes (MDP) \cite{zhang2021multi}. The goal is typically to learn an optimal load balancing policy, which has been done by using numerous available multi-agent reinforcement learning (MARL) approaches, mainly building upon multi-agent MDPs (MA-MDP) \cite{chu2020multi, lin2021multi} and decentralized partially observable MDPs (Dec-POMDP) \cite{guicheng2022review}. 
However, these methods are usually either not highly scalable or are difficult to analyze \cite{zhou2021asymptotically}. This has resulted in recent popularity of mean field theory for modelling multi-agent systems, via a single representative agent interacting with the mean field (distribution) of all agents \cite{lacker2019local, cui2022survey}. Within mean field theory, one differentiates between mean field games (MFG) for agents in a competitive setting \cite{lasry2007mean} and mean field control (MFC) for cooperation \cite{bensoussan2013mean}. 
This work focuses on the latter for reducing the overall jobs dropped in the system. 

Note that mean field limits have also been used to study load balancing algorithms, such as JSQ and the power-of-$d$ variants (where the load balancer only obtains the state information of $d \ll M$ out of the total available queues) \cite{mitzenmacher2001power}, in terms of sojourn time and average queue lengths \cite{ dawson2005balancing}. 
The asymptotic analysis of load balancing policies for systems with different underlying random dense graphs has already been studied to show that as long as the degree $d(M)$ scales with the number of servers, the topology does not affect the performance \cite{mukherjee2018asymptotically}, while in our work we consider the sparse case.
Graphons are also used to describe the limit of dense graph sequences \cite{lovasz2012large} and have been studied with respect to both MFG and MFC \cite{hu2022graphon, cui2021learning}. 
However, for sparse graphs, the limiting graphon is not meaningful, making it unsuitable for networks whose degree does not scale with system size, see also \cite{caines2021graphon, lovasz2012large}. 
MFGs for relatively sparse networks have been studied using Lp-graphs \cite{fabian2022learning}, but they too cannot be extended to bounded degrees.
We believe that bounded-degree graph modelling is necessary to truly represent and analyse large fixed-degree distributed systems to avoid the need for increasingly global interactions and scaling difficulties.

In models where the graph degree is fixed and small, for different topologies such as the ring and torus, the power-of-two policy was studied using pair approximation to analyse its steady state and to show that choosing between the shorter of two local neighbouring queues improves the performance drastically over purely random job allocations \cite{gast2015power}.
In contrast, in this work, we will also look at a similar model, where the degree $d$ does not scale with the number of agents, and find that one can improve beyond existing power-of-$d$ policies. In order to find scalable load balancing policies, we apply relatively new results from sparse mean field approximations, see \cite{lacker2019local} and references therein.
Recently, mean field analysis of the load balancing policies was done using coupling approach but mainly for spatial graphs and only for power-of-$d$ algorithms, see \cite{rutten2023mean}. They consider tasks and servers which may be non-exchangeable on random bipartite geometric graphs and random regular bipartite graphs. 
To the best of our knowledge, MFC on sparse (bounded degree) graphs has neither been analyzed theoretically in its general form, nor has it been applied to queueing systems. 

Additionally, algorithms such as JSQ, SED and their power-of-$d$ variant are based on the unrealistic assumption of instantaneous knowledge of the queue states,  which is not true in practice, especially in large systems.
We remove this assumption by introducing a synchronization (communication) delay $\Delta t$ into our system model. 
Though in non-local systems it has already been proven that for $\Delta t=0$, JSQ is the optimal load balancing policy, it is also known that as the delay increases, JSQ fails due to the phenomenon of "herd behaviour" \cite{mitzenmacher2001power}, i.e. allocating all packets simultaneously to momentarily empty queues.
It has also been shown that as $\Delta t \to \infty$, random allocation is the optimal policy \cite{mitzenmacher2001power}. While for the in-between range of synchronization delays in non-sparse queueing systems, a scalable policy has been learned before using mean field approximations \cite{tahir2022learning}, it only considered fully connected graphs with sampling $d$ neighbours, and not the sometimes more realistic scenario of local, sparsely connected queueing networks. For other works in this direction, we refer to \cite{mukherjee2018asymptotically, zhou2021asymptotically, lipshutz2019open} and references therein.

\paragraph*{Contributions}
Our main contributions through this work are: (i) we model a novel MFC problem with an underlying sparse bounded-degree graph, giving a first tractable and rigorous description of general large distributed queueing systems with \textit{sparse interconnections}; (ii) we provide theoretical approximation guarantees to motivate our model; (iii) we consider a communication delay in distributed queueing systems, so that all agents can work in a synchronized manner resulting in a discrete-time model; and (iv) we use state-of-the-art reinforcement learning (RL) algorithms to successfully find scalable load balancing algorithms for distributed systems.

\section{Queueing System}
\label{sec:mathematical_model}
We consider a system with a set of $\mathcal N = \{1, \ldots, N\}$ schedulers/agents and $\mathcal M = \{1, \ldots, M\}$ servers, where each server has its own queue with finite buffer capacity, $B$.
The queues work in a first-in-first-out (FIFO) manner, and servers take the jobs one at a time from their queue, processing them at an exponential rate $\alpha$. Once a job has been processed, it leaves the system forever.
Somewhat to the successful power-of-$d$ policies \cite{mitzenmacher2001power}, we assume that each scheduler accesses only a limited number (e.g. $d$ out of the $M$) of available queues and can only allocate its arriving jobs to these $d$ accessible queues, with $d \ll M$ and fixed.
We assume $M=N$ and associate each server (queue) with one agent, though it is possible to extend the model to varying or even random numbers of queues per agent.
Note that all connections between agents are assumed to be wireless.

Jobs arrive to the system accordingly to a Markov modulated Poisson process \cite{fischer1993markov} with total rate $\lambda(t) N$, and then are divided uniformly amongst all agents, which is also equivalent to independent Poisson processes at each agent given the shared arrival rate $\lambda(t)$ by Poisson thinning \cite{harremoes2007thinning}. 
The agent takes the allocation action based on a learned or predetermined, memory-less policy $\zeta$, which considers current queue state information of its $d$ accessible queues. 
This information is periodically sent by servers to neighbouring agents, such that agents only obtain information on $d$ queues, reducing the amount of messages to be sent.
If the job allocation is done to an already full queue, it is dropped and results in a penalty. 
Similarly, jobs depart from a queue at rate $\alpha$.
The goal of the agent is to minimize the overall jobs dropped.


\vspace{-0.2cm}
\subsection{Locality and Scalability}
Note that in contrast to many other analyzed settings, in this work the $d$ out of $M$ available queues are not sampled randomly for each package, but instead fixed for each agent according to some predefined topology (see also Section~\ref{sec:topologies}). In other words, we assume a strong concept of locality where agents have access only to a very limited subset of queues, implying also a strong sense of partial observability in the system.
The value for $d$ therefore depends on the type of graph topology being considered.
Note that an agent always has access to its own queue.
Accordingly, our queueing model contains an associated static underlying 
undirected graph $G = (\mathcal N, \mathcal E)$, where $\mathcal E \subseteq \mathcal N \times \mathcal N$ is the set of wireless edges between vertices (agents) $\mathcal N \coloneqq \{1, \ldots, N\}$ based on the $d$ queues accessible by the agent.
An agent $i$ will have an edge $(i,j) \in \mathcal E$ to another agent $j$ whenever they have access to the queues associated to that agent, and vice versa. Therefore, an agent can have a maximum of $d$ edges (neighbours) to other agents. 
We will denote the set of neighbours of agent $i$ as $N_i \coloneqq \{ j \in \mathcal N \mid (i,j) \in \mathcal E\}$.

The motivation to use a graph structure with bounded degree arises from the fact that the corresponding model allows us to find more tractable local solutions for large systems. We need large systems that are structured in some sense, otherwise the system is too heterogeneous to find a tractable and sufficiently structured solution. Therefore, we look at systems where the structure can be expressed graphically. To avoid confusion, note that the framework is not limited to regular graphs. The framework already includes basic regular graphs such as grids and toruses, but also allows many other irregular random graphs such as the configuration model, see also Section \ref{sec:topologies}. Here we then apply RL and MFC to find otherwise hard-to-compute near-optimal queueing strategies in this highly local setting. The simplicity of the queueing strategy -- instantaneously assigning a packet to one of the neighbouring queues based on periodic and local information -- not only allows for fast execution and high scalability, since information does not need to be exchanged for each incoming packet, but also allows for easy addition of more nodes to scale the system to arbitrary sizes.
So in the following, we will obtain tractable solutions by first formulating the finite queueing system, then formulating its limiting mean field equivalent as the system grows, and lastly applying (partially-observed) RL and MFC.

\subsection{Finite Queueing System}
To begin, consider the following system. Each agent $i$ is associated with a local state, $z_i \in \mathcal Z \coloneqq \{0, 1, \ldots, B\}$ and local action $a_i \in \mathcal A_i$. The state $z_i$ will be the current queue filling of the $d$ queues from $\mathcal M$ accessible to agent $i$. The set of actions $\mathcal A_i$ could be the set of these $d$ accessible queues, to which new packets are assigned. Hence, we have a finite set of action and state space.
The global state of the system is given by $\mathbf z = (z_1, \ldots, z_N) \in \mathcal Z^N$. Similarly, the global action is defined as $\mathbf a = (a_1, \ldots, a_N) \in \mathcal A^N \coloneqq \mathcal A_1 \times \cdots \times \mathcal A_N$. 

\subsubsection{Synchronous model} 
We want the agents to work in a synchronized manner, e.g. to model wireless communication delays, and also for the servers to send their queue state information to the respective agents once every fixed time interval. To achieve this, we model our system at discrete decision epochs $\{0, \Delta t, 2 \Delta t, 3 \Delta t, \ldots\}$, where $\Delta t > 0$ is the synchronization delay, the time passing between each decision epoch. The interval $\Delta t$ may be understood as a type of synchronization or update delay, assuming that it takes $\Delta t$ amount of time to obtain updated local information from the servers and update the queueing strategy (e.g. routing table). Note that we may easily adjust $\Delta t$ to approximate continuous time. Using this delay, we can also model our system as a MFC-based RL problem and learn the optimal policy using state-of-the-art RL-based algorithms \cite{sutton2018reinforcement}.

\subsubsection{Localized policies} 
For the moment, let each agent be associated with a localized policy $\zeta_i^{\theta_i}$ parameterized by $\theta_i$ which defines a distribution of the local action $a_i$ conditioned on the states of its neighbourhood $\mathbf z_{N_i}(t)$.
The size of the neighbourhood depends on $d$ and the type of graph used.
Given the global state $\mathbf z(t)$, each agent takes an action $a_i(t)$ which is independently drawn from $\zeta_i^{\theta_i}(\cdot \mid z_{N_i}(t))$.
In other words, agents do not coordinate between each other, and decide independently where to send arriving packets.
The parameters $\theta = (\theta_1, \ldots, \theta_N)$ parametrize the tuple of localized policies $\zeta_i^{\theta_i}$, with the joint policy as its product $\zeta^{\theta}(\mathbf a \mid \mathbf z) = \prod_{i=1}^N \zeta_i^{\theta_i}(a_i \mid z_{N_i})$, as agents act independently.

\textit{Note: We do not include the actions of other agents, because all agents may take an action at the same time, so we will not have this information and each agent will act independently.}

\subsubsection{Symmetrized packet assignments} 
For tractable analysis, it is necessary to handle neighbors anonymously, i.e., providing the same probability of assigning to any particular neighbors with the same queue states.
Indeed, to some extent such symmetrization is necessary to obtain a useful mean field limit. Otherwise, behaviour depends strongly on the ordering of neighbours, which we cannot model to obtain the mean field limit. 
Consider a simple one-dimensional line graph where nodes are connected in a straight path. If we allow actions that are not symmetric, e.g. if all agents send all their packets to the first of their two neighbours under some ordering of their neighbours, then we obtain different behaviour depending on this ordering. Cutting the graph into sets of three nodes, we could define the center node of each set to be the first neighbour for the other nodes, leading to a packet arrival rate of $2 \lambda(t)$ at the center node. On the other hand, if we define first neighbours such that there are no overlapping first neighbours between any agents, then any node will have a packet arrival rate of at most $\lambda(t)$. Hence, a certain symmetrization is natural for the scaling limit.

\begin{figure}
\centering
\includegraphics[width=0.65\linewidth]{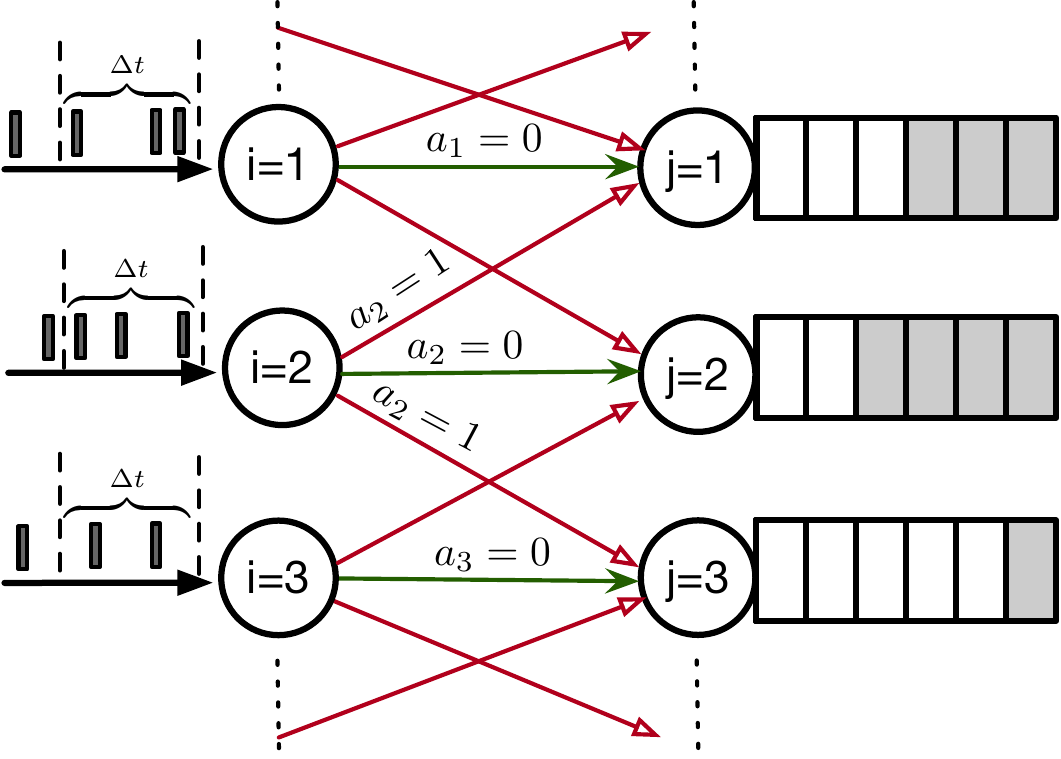}
\caption{Visualization of how agents implement their policy. For instance, agent $i=2$ has neighbours $j\in\{1,3\}$. If its action is $a_2=0$, it will allocate all its arriving jobs to its own queue $j=2$ (green arrow). In contrast, if $a_2=1$ then the arriving jobs are allocated randomly to one of its neighbour $j$ (red arrows).
}
\label{fig:policy_visualisation}
\end{figure}

At most, we could consider a solution that anonymizes but still differentiates between neighbors whenever one is fuller than the other. This may be important especially if e.g. queue serving rates are heterogeneous, and the according framework is straightforward. However, in our experiments we found that such an assumption complicates training of a RL policy due to the significant addition of action space complexity. Therefore, trading off between policy expressiveness and RL training stability, we assume that any agent may choose to either send to their own queue, or offload uniformly randomly to any of their neighbours. In other words, for simplicity of learning we consider the actions $\mathcal A_i = \mathcal A = \{0, 1\}$ for all agents $i$, of either sending to its own queue ($0$), or randomly sending to a neighbour ($1$), see Figure \ref{fig:policy_visualisation} for an example visualization. This assumption symmetrizes the model and obtains better performance, see Appendix.

\subsubsection{Dynamical system model} 
The state of the agent is the state of its associated queue, and is affected by the actions of itself and neighbouring agents. By Poisson thinning, at every epoch $t \in \mathbb N$, given the current global state $\mathbf z(t)$ and action $\mathbf a(t)$, the next local state $z_i(t+1)$ of agent $i$ can be calculated independently only depending on the current state $\mathbf z_{i}(t)$ and actions $\mathbf a_{N_i}(t) \equiv \{a_j(t)\}_{j \in N_i}$, i.e. $    \mathbb P(\mathbf z(t+1) \mid \mathbf z(t), \mathbf a(t)) = \prod_{i=1}^N \mathbb P(z_i(t+1) \mid z_{i}(t), a_{N_i}(t))$,
where each $\mathbb P(z_i(t+1) \mid z_{i}(t), a_{N_i}(t))$ can be computed by the Kolmogorov equation for continuous-time Markov chains, given that the rate of an arrival at a queue is given by the sum of arrival rates assigned by the agents, $\lambda_t^i = \lambda(t) (1 - a_i + \sum_{j \in N_i} \frac{a_j}{|N_j|})$ such that the total arrival rate is $\sum_i \lambda_t^i = N\lambda(t)$, and the package departure rate of each queue is fixed to the serving rate $\alpha > 0$.


Each agent is associated with a local stage reward $r_i(z_i,a_i)$ and according global stage reward $r(\mathbf z, \mathbf a) = \frac 1 N \sum_{i=1}^N r_i(z_i,a_i)$. The reward is given in terms of a penalty for \textbf{job drops} due to each action $a_i$.
The objective is then to find the localized policy tuple $\theta$ such that the global reward is maximized, starting from initial state distribution $z_i(0) \sim \mu_0$.

\subsection{Sparse Graph Mean Field System}
In this section, we will consider limits of large graphs and the behaviour in such systems, by using mean field theory from \cite{lacker2019local} which focuses on an abstract theoretical framework, that is (i) unrelated to particular applications and (ii) does not explicitly consider control, which we integrate into the states of their discrete-time model.

In order to obtain a limiting mean field system, we assume a shared policy $\zeta_i = \zeta$ for all agents $i$. This assumption is natural, as it often gives state-of-the-art performance in various RL benchmarks \cite{yu2021surprising, christianos2020shared, papoudakis2021benchmarking} and also allows immediate application to arbitrary sizes of the finite system, as all schedulers can use the same policy. Furthermore, it will allow to scale up the system at any time by simply adding more schedulers and queues, without retraining a policy.

In contrast to typical mean field games \cite{lasry2007mean} and mean field control \cite{bensoussan2013mean}, we cannot reduce the entire system exactly to a single representative agent and its probability law. This is because in a sparse setting, the neighbourhood state distribution of any agent remains an empirical distribution of finitely many neighbours and hence cannot follow a law of large numbers into a deterministic mean field distribution. Therefore, the neighbourhood and its state distribution remain stochastic and it is necessary to model the probability law of entire neighbourhoods. The modelling of such graphical neighbourhoods is formalized in the following.

\subsubsection{Topological structure} 
To make sense of the limiting topology of our system formally, we introduce technical details, letting finite systems be given by sequences of (potentially random) finite graphs and initial states $(G_n, \mathbf z_n)$ converging in probability in the local weak sense to some limiting random graph $(G, \mathbf z)$. Here, we assume that graphs are of bounded degree, i.e. there exists a finite degree $d$ such that all nodes have at most $d$ neighbours. In other words, we define a sequence of systems of increasing size according to a certain topology, which formalizes the scalable architectural choice of a network structure, such as a ring topology or torus. In the following paragraph, we give the formal definition, which may be skimmed by the reader.

For convergence in the local weak sense, define first the space of marked rooted graphs $\mathcal G_*$, the elements of which essentially constitute a snapshot of the entire system at any point in time. Such a marked rooted graph consists of a tuple $(G, \varnothing, \mathbf z) \in \mathcal G_*$, where $G = (\mathcal N, \mathcal E)$ is a graph, $\varnothing \in \mathcal N$ is a particular node of $G$ (the so-called root node), and $z \in \mathcal Z^{\mathcal N}$ defines states ("marks") for each node in $G$, i.e. the current queue filling of queues associated to any agent (node). Denote by $B_k(G, \varnothing, \mathbf z)$ the marked rooted subgraph of vertices in the $k$-hop neighbourhood of the root node $\varnothing$. The space $\mathcal G_*$ is metrized such that sequences $(G_n, \varnothing_n, \mathbf z_n) \to (G, \varnothing, \mathbf z) \in \mathcal G_*$ whenever for any $k \in \mathbb N$, there exists $n'$ such that for all $n > n'$ there exists a mark-preserving isomorphism $\phi \colon B_k(G_n, \varnothing_n, \mathbf z_n) \to B_k(G, \varnothing, \mathbf z)$, i.e. with $z_{ni} = z_{(\phi(i))}$ for all nodes $i \in G_n$ (local convergence).
We will abbreviate elements $(G_n, \varnothing_n, \mathbf z_n), (G, \varnothing, \mathbf z) \in \mathcal G_*$ as $(G_n, \mathbf z_n), (G, \mathbf z)$, and their node sets as $G_n, G$ whenever it is clear from the context. Then, finally, convergence in the local weak sense is formally defined by 
$    \lim_{n \to \infty} \frac{1}{|G_n|} \sum_{i \in G_n} f(C_i(G_n)) = \mathbb E \left[ f(G) \right]$
in probability for every continuous and bounded $f \colon \mathcal G_* \to \mathbb R$, where $C_i(G_n)$ denotes the connected component of $i$ in $G_n$. 

In other words, wherever we randomly look in the graph, there will be little difference between the distribution of the random local system state (including its topology), and of the limiting $G$. This holds true, e.g. if we initialize all queues as empty or i.i.d., and consider certain types of topologies such as grids, see Section~\ref{sec:topologies}. More details also in \cite{lacker2019local}.

\subsubsection{System model} 
As discussed in the prequel, consider sequences of possibly random rooted marked graphs (i.e. finite graphs and initial states) $(G_n, \mathbf z_n) \to (G, \mathbf z)$ with $|G_n|$ agents $i=\{1,\ldots,|G_n|\}$, converging in the local weak sense to the potentially infinite-sized system $(G, \mathbf z)$. For a moment, consider a decentralized, stochastic control policy $\zeta \colon \mathcal Z \times \mathcal T \to [0, 1]$ such that an agent $i \in G$ chooses to offload to a random neighbour (action $a_i=1$) with probability $\zeta_t(z_i(t))$, depending on its own queue state $z_i(t)$ only. We can then consider the probability law $\mathcal L(G, \mathbf z)$ of the limiting system $(G, \mathbf z)$ as the state of a single-agent MDP, similar to the MFC MDP formalism in standard MFC \cite{pham2018bellman}. This law is in essence the probability that the graph around any randomly chosen queue is in a certain state. At least formally, we do so by identifying the choice of $\zeta_t$ at any time $t \in \mathbb N$ as an action, and letting the state of the MFC MDP be given by the law of the time-variant system $(G(t), \mathbf z(t))$ resulting from the application of $\zeta_t$ when starting at some initial law.

Deferring for a moment (until Section~\ref{sec:rl}) (i) the question of how to simulate, represent or compute the probability law of a possibly infinite-sized rooted marked graph, and (ii) the detailed partial information structure (i.e. observation inputs) of the following policy, we consider a hierarchical upper-level MFC policy $\pi$ that, for any current MFC MDP state, assigns to all agents at once such a policy $\zeta_t = \pi(\mathcal L(G(t), \mathbf z(t)))$ at time $t$. To reiterate, the decentralized control policy $\zeta_t$ at any time $t$ now becomes the action of the MFC MDP, where the dynamics are formally given by the usually infinite-dimensional states $\mathcal L(G(t), \mathbf z(t))$, i.e. the probabilities of the limiting rooted marked graph $(G(t), \mathbf z(t))$ being in a particular state after applying a sequence of policies $(\zeta_0, \ldots, \zeta_{t-1})$. The cost function $J^{G, \mathbf z}(\pi)$ for any such upper-level policy $\pi$ is then given by the number of expected packet drops per agent $D(t)$ in the limiting system,
\begin{align*}
    J^{G, \mathbf z}(\pi)
    &= - \mathbb E \left[\sum_{t=0}^{\infty}\gamma^t D(t) \right].
\end{align*}

Using analogous definitions for the finite system based on the topologies and initial states $(G_n, \mathbf z_n)$, we can apply the upper-level MFC policy $\pi$ to each agent and use the resulting number of average packet drops $D^N(t)$ at time $t$. Using our newly introduced graphical formulation, we thus write the cost function $J(\theta)$ in the finite system $(G_n, \mathbf z_n)$ as
\begin{align*}
    J^{G_n, \mathbf z_n}(\theta)
    &= - \mathbb E \left[\sum_{t=0}^{\infty}\gamma^t D^N(t) \right].
\end{align*}

\subsubsection{Optimality guarantees} 
One can now show that the performance of the finite system is approximately guaranteed by the performance in the limiting mean field system. Informally, this means that for any two policies, if the performance of one policy is better in the mean field system, it will also be better in large finite systems.

\begin{theorem} \label{thm:Jconv}
Consider a sequence of finite graphs and initial states $(G_n, \mathbf z_n)$ converging in probability in the local weak sense to some limiting $(G, \mathbf z)$. For any policy $\pi$, as $n \to \infty$, we have convergence of the expected packet drop objective
\begin{align*}
    J^{G_n, \mathbf z_n}(\pi) \to J^{G, \mathbf z}(\pi).
\end{align*}
\end{theorem}

As a result, we have obtained a limiting mean field system for large systems, which may be more tractable for finding improved load balancing schemes. In particular, if we have a number of policies between which to choose, then the policy performing best in the MFC system will also perform best in sufficiently large systems. Here, the set of MFC policies can include well-known algorithms such as JSQ, which is known to be optimal in many special cases.

\begin{corollary} \label{coro:kopt}
    Consider a finite set $\{\pi_1, \ldots, \pi_K\}$ of MFC policies with differing MFC objective values $J^{G, \mathbf z}(\pi_1)$ to $J^{G, \mathbf z}(\pi_K)$. Let $\pi_i$ be the policy with maximal MFC objective value $J^{G, \mathbf z}(\pi_i) \geq \max_j J^{G, \mathbf z}(\pi_j)$. Then, there exists $n'$ such that for all $n > n'$, we also have optimality in the finite system
    \begin{align*}
        J^{G_n, \mathbf z_n}(\pi_i) \geq \max_j J^{G_n, \mathbf z_n}(\pi_j).
    \end{align*}
\end{corollary}

The proofs are provided in Appendix and use the theoretical framework of \cite{lacker2019local} for general dynamical systems. 
In our experiments, we will also allow for randomized assignments per packet to further improve performance empirically, such that all packets arriving at a particular scheduler during the entirety of any epoch are independently randomly either allocated to the local queue or a random neighbour, i.e. formally we replace offloading choices $\mathcal A = \{0, 1\}$ by probabilities for offloading each packet independently, $\mathcal A = [0, 1]$.

The MFC MDP formulation and theoretical guarantees give us the opportunity to use MFC together with single-agent control such as RL in order to find good scalable solutions while circumventing hard exact analysis and improving over powerful techniques such as JSQ in large queueing systems. All that remains is to solve the limiting MDP, as MFC has formally converted MARL into single-agent RL. 


\subsection{Reinforcement Learning}
\label{sec:rl}

Building upon the preceding MFC formulation, all that remains to find optimal load balancing in large sparse graphs, is to solve the MFC problem. Due to its complexity, the limiting MFC problem will be solved by considering it as a variant of an MDP, i.e. a standard formulation for single-agent centralized RL, which will also allow a model-free design methodology. Nonetheless, the training is still performed on the preceding finite graph multi-agent system, as we cannot evaluate the typically infinite limiting system $\mathcal L(G(t), \mathbf z(t))$. 

More precisely, we will consider a partially-observed MDP (POMDP) variant of the problem, since at any time $t$, we cannot evaluate the potentially infinite system $\mathcal L(G(t), \mathbf z(t))$ exactly to obtain action $\zeta_t = \pi(\mathcal L(G(t), \mathbf z(t)))$. Instead, we will use the empirical distribution $\mu^{G_n, \mathbf z_n}(t)$ as an observation that is only correlated with the state of the entire system, but of significantly lower dimensionality ($|\mathcal Z|$-dimensional vector, instead of $\mathcal Z^{|G|}$ plus additional topological information). This also means that we need not consider the limiting system of potentially infinite size, or include the information of root nodes when considering network topologies in the following, which is intuitive as there is no notion of global root in local queueing systems.

Here, the centralized RL controller could have estimated or exact global information on the statistics of the queue states of all nodes, or alternatively we can understand the approach as an optimal open-loop solution for any given known starting state, since the limiting MFC dynamics on $\mu^{G_n, \mathbf z_n}(t)$ are deterministic, and therefore the centralized RL controller can be used to compute an optimal deterministic sequence of control, which can then be applied locally.

In our experiments we also allow to simply insert the empirical distribution of the locally observed neighbour queue states at each node (a simple estimate of the true empirical distribution),
to instead sample decision rules for each agent according to the local empirical state distribution, which is verified to be successful. Thus, our approach leans into the centralized training decentralized execution scheme \cite{zhang2021multi} and learns a centralized policy, which can then be executed in a decentralized manner among all schedulers. As desired, our approach is applicable to localized queueing systems.

In order to solve the POMDP, we apply the established proximal policy optimization (PPO) RL method \cite{schulman2017proximal, yu2021surprising} with and without recurrent policies, as commonly and successfully used in POMDP problems \cite{ni2021recurrent}. PPO is a policy gradient method with a clipping term in the loss function, such that the policy does not take gradient steps that are too large while learning \cite{schulman2017proximal, yu2021surprising}. The overall training algorithm is given in Appendix, which also shows how to analogously apply a trained MFC policy to a finite system.

\subsubsection{Training on a finite queueing system} 
Note that the considered observation $\mu^{G_n, \mathbf z_n}(t)$ and any other variables such as the number of dropped packets at the root node can indeed be computed without evaluating an infinite system until any finite time $t$, since at any time $t$, at most any node less than $t$ steps away from the root node may have had an effect on the root node state. Therefore, the computation of root node marginals can be performed exactly until any finite time $t$, even if the limiting system consists of an infinitely large graph $G$. However, the cost of such an approach would still be exponential in the number of time steps, as a $k$-hop neighbourhood would typically include exponentially many nodes, except for very simple graphs with degree $2$. 

We therefore consider alternatives: For one, we could apply a sequential Monte Carlo approach to the problem by simulating $M$ instances of a system from times $0$ to some terminal time $T$ that consists of all nodes less than $T$ away from the root node. However, this means that we would have to simulate many finite systems in parallel. Instead, using the fact that the empirical distribution of agent states $\mu^{G_n, \mathbf z_n}(t)$ in the finite system converges to $\mathcal L(X^{G,x}_{\varnothing}(t))$ as seen in Theorem~\ref{thm:Jconv}, it should be sufficient to evaluate $\mathcal L(X^{G,x}_{\varnothing}(t))$ via the empirical distribution of a sufficiently large system. 

Thus, we simulate only a single instance of a large system with many nodes by using it for the limiting MFC, which is equivalent to learning directly on a large finite system. In other words, our approach learns load balancing strategies on a finite system by using the MFC formalism for tractability of state and action representations, with theoretical guarantees.


\section{Experiment Setup}

In this section, we give an explanation of the different types of graph topologies we have used to verify our aforementioned theoretical analysis. We also give a description of the different used load balancing algorithms.

\subsection{Topologies}
\label{sec:topologies}
A brief description of the practical topologies of interest, most of which fulfill the convergence in the local weak sense defined earlier is given here, where the agents are numbered from $1$ to $N$.
(i) First, we consider the simple CYC-1D graph, which has extensively been used in the study of queueing networks and is highly local \cite{shortle2018fundamentals}.  Each agent $i$ has access to $d=2$ other queues/servers, $\{i-1, i+1\}$, while the edge nodes, $1$ and $N$, form a connection. 
(ii) Next, we define the \textbf{cube-connected cycle} (CCC) graph. This undirected cubic graph has been considered as a network topology for parallel computing architectures \cite{ habibian2017fault}. It is characterized by the cycle order, $o_c$ which is the degree $d$ of each node and defines the total number of nodes $N = o_c2^{o_c}$ in the graph. 
(iii) We also apply the \textbf{torus} (TORUS) grid graph that has been repeatedly used to represent distributed systems for parallel computing \cite{mehmet2019networks}, 
as a higher-dimensional extension of the CYC-1D graph. We here consider a $2$-D torus which is a rectangular lattice having $o_t$ rows and columns. Every node connects to its $d=4$ nearest neighbours and the corresponding edge nodes are also connected. The total nodes in a $2$-D torus are $N=o_t^2$.
(iv) Another highly general topology is the \textbf{configuration model} (CM). This sophisticated generalized random graph is one of the most important theoretical models in the study of large networks \cite{fosdick2018configuring}. 
In contrast to the prior highly local topologies, the CM can capture realistic degree distributions under little clustering. Here every agent is assigned a certain degree, making the graph heterogeneous as compared to every agent having the same degree as in previous mentioned graphs. The degree sequence we have used is in the set $\mathcal D = \{2,3\}$ with equal likelihood.
(v) And lastly, for an ablation study on graphical convergence assumptions, we use the \textbf{Bethe} (BETHE) lattice. This cycle-free regular tree graph is used for analysis of many statistical physics, mathematics related models and potential games \cite{szabo2016evolutionary}. 
It is characterized by a pre-defined lattice depth $o_b$, with all nodes in the lattice having the same fixed number of neighbours, $d$. The number of nodes at a depth $o$ away from the root node are given by: $d(d-1)^{(o_b-1)}$ and the total nodes in the Bethe lattice are calculated as: $N = 1 + \sum_{o=1}^{o_b} d(d-1)^{(o-1)}$. We use $d=3$ for our experiments.
Note that it has already been formally shown in \cite[Sections 3.6 and 7.3]{lacker2019local} that for a sequence of increasing regular trees, the empirical measure does not converge to the same limit law as the root particle, even in the weak sense. 
This is because even as $N \to \infty$, a large proportion of the nodes are leaves, which greatly influences the behaviour of the empirical measure, as particles at different heights behave differently, and the root particle is the only particle at height zero. We have also verified this mismatch using our experiments in Section~\ref{sect:experiments}, though we nevertheless obtain improved performance in certain regimes. 


\subsection{Load Balancing Algorithms}
\label{sec:solvers}
We now explain first the mean field solver that we have designed, based on our mathematical modelling from Section \ref{sec:mathematical_model}, namely the MF-Random solver. Then we mention all the existing state-of-the-art finite-agent solvers which we have used for performance comparison. The different kinds of load balancing policies which we have verified on the above-mentioned topologies are given in the following:
\begin{itemize}
    \item MF-Random (MF-R): The agent is only aware of the state of its own queue. The upper-level learned policy is a vector that gives the probability of sending jobs to the own queue or to a random other accessible queue.
    \item MARL-PS (NA-PS): The policy is trained, using PPO with parameter sharing \cite{ christianos2020shared}, as it often gives state-of-the-art performance in various RL benchmarks \cite{yu2021surprising, papoudakis2021benchmarking}, on a smaller number of agents. The learned policy can be used for any arbitrary number of agents. It generates a continuous policy which gives the probability of either sending to your queue or randomly to one of the neighbours queues while only observing the state of your own queue, similar to MF-R.
    \item Join-the-shortest-queue (JSQ): A discrete policy that sends jobs to the shortest out of all accessible queues.
    \item Random (RND): A policy where the probability of sending the job to any accessible queues is equally likely.
    \item Send-to-own-queue (OWN): A discrete policy where the agent only sends jobs to its own queue.
\end{itemize}

Note that the action from the above-mentioned load balancing policies is obtained at the beginning of each $\Delta t$ and then used for that entire $\Delta t$ timestep. Also note that for all the algorithms, the value for the number of accessible queues $d$ is dependent on the type of graph topology being used, and we refer to Section \ref{sec:topologies} for more details on this. In the next section, we discuss all the required parameters and chosen values for our experiments.




\begin{figure*}
\centering
\includegraphics[width=\linewidth]{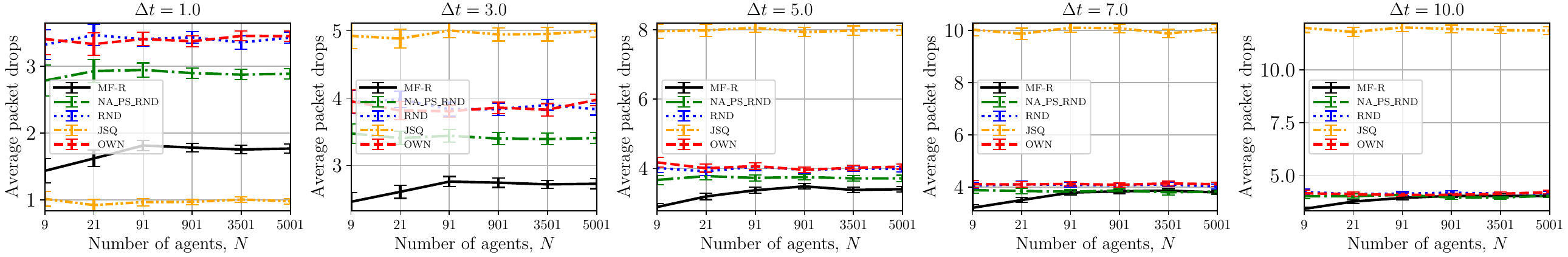}
\caption{Performance comparison of the learned MF-R policy to NA-PS, JSQ, RND and OWN algorithms, on a CYC-1D graph, 
over a range of $\Delta t$s
is shown, with $95\%$ confidence intervals depicted by error bars. The degree of each agent is $d=2$ and the number of agents (queues) used to make the graph are $N\in\{9, 21, 91, 901, 3501, 5001 \}$. 
}
\label{fig:cyc1d}
\end{figure*}


\subsection{Training}
For all our experiments, we consider that each agent is associated with only one queue, so $N=M$. The servers work at an exponential rate of $\alpha$. At every decision epoch we simulate a Markov modulated arrival rate $\lambda_{t}$, with $\lambda_0 \sim \mathrm{Unif}(\{\lambda_h, \lambda_l\})$ with transition law for switching between rates: $\mathbb P(\lambda_{t+1} = \lambda_l \mid \lambda_t = \lambda_h) = 0.2,$ and $\mathbb P(\lambda_{t+1} = \lambda_h \mid \lambda_t = \lambda_l) = 0.5$.
Note that these values were chosen to depict the switching of the system between high and low traffic regimes. In principle, any reasonable values can be considered. 
We use the well-established policy gradient based RL algorithm, proximal policy optimization (PPO) with a localized reward function which penalizes for the drops in each queue.
The rest of the system parameters and the PPO hyperparameters are given in Appendix.
Lastly, the number of agents $N$ and degrees $d$ in different graph topologies are fixed during training of the MF-R policies as:
\begin{itemize}
    \item $1$-D cyclic (CYC-1D): $N=M=101$, $d=2$,
    \item Cube-connected cycle (CCC): $o_c=5$, $N=125$, $d=3$,
    \item $2$-D Torus grid (TORUS): $o_t=11$, $N=111$, $d=4$,
    \item Configuration model (CM): $N=101$, $d\in\{2,3\}$,
    \item Bethe lattice (BETHE): $o_b=5$, $N=94$, $d=3$.
\end{itemize}
Once trained on these parameters, the learned policy can be evaluated on varying graph sizes without the need of retraining, as done in our experiments.


\section{Experiment Results}
\label{sect:experiments}

We now present the performance comparison of the load balancing policies of Section~\ref{sec:solvers} on graph topologies from Section~\ref{sec:topologies}. 
For the exact simulation of associated continuous-time Markov chains $y$, we sample exponential waiting times of all events using the Gillespie algorithm.
For training, we use the same simulation horizon for each episode consisting of $T = 50$ discrete decision epochs, and for comparability the performance is evaluated in terms of the average packet drops which is calculated as the sum over $T$ decision epochs of the average number of total packets dropped in all queues. However, note that simulating the same time span with different $\Delta t$ does provide slightly different results, due to the switching between high and low traffic regimes after each $\Delta t$ epoch. Each evaluation was repeated for $100$ simulated episodes, and error bars depict the $95 \%$ confidence interval.

\paragraph{Finite system performance}
The second result of interest is the performance analysis on the CYC-1D graph over a range of $\Delta t$ for the learned MF-R policy as compared to the NA-PS, JSQ, RND and OWN algorithms, see Figure \ref{fig:cyc1d}. The number of agents (queues) used to generate the graph are $N \in \{9, 21, 91, 901, 3501, 5001 \}$, with every agent always having the fixed degree $d=2$. For this graph, we also trained an NA-PS policy on a sufficiently small size with $N=M=5$ to obtain convergence even with batch and minibatch sizes of $50000$ and $8000$. We used PPO with parameter sharing so that the learned policy could be evaluated on any graph size. 

\begin{figure}
\centering
\includegraphics[width=0.5\linewidth]{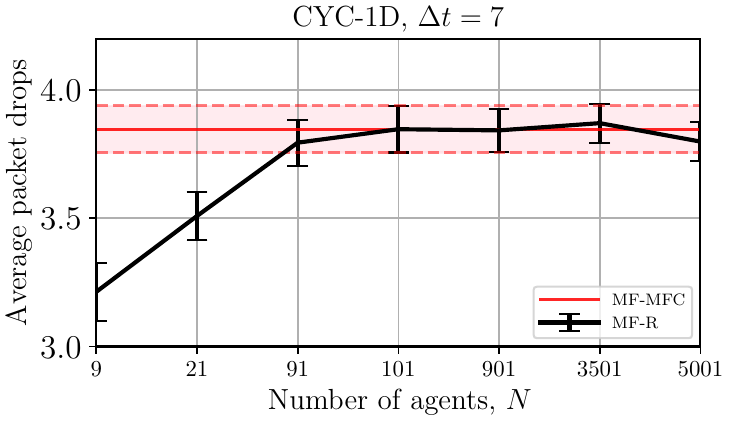}
\caption{Performance of the MF-R policy for increasingly large CYC-1D graphs. The red horizontal line indicates the evaluated episode return of the learned MF-R policy during training on $N=101$, (MF-MFC). Shaded regions depict the $95\%$ confidence intervals. 
}
\label{fig:mf_plot}
\end{figure}

It can be seen that JSQ is the best strategy at $\Delta t=1$, while the performance of MF-R is very close to JSQ. And as expected, the performance of JSQ deteriorates with increasing synchronisation delay $\Delta t$. For an intermediate value of 
$\Delta t \in \{3, 5, 7\}$, 
neither JSQ nor RND are the optimal load balancing policies. Here, our learned MF-R policy performs best and approaches the optimal RND performance as $\Delta t$ increases to $10$. As discussed earlier, it has already been proven in certain scenarios that as $\Delta t \to \infty$, RND is the optimal load balancing policy \cite{mitzenmacher2001power}.
The learned NA-PS policy has a similar trend to our MF-R policy, but is unable to outperform it even for small systems.
For the CYC-1D, 
CCC, and TORUS graphs with homogeneous degrees, the performance of OWN in the limit -- sending only to its own queue -- is equivalent to RND, since for homogeneous degrees, sending to your own queue results in the same packet arrival rate at each queue as randomly sending to any queue in your neighbourhood.
In Figure \ref{fig:mf_plot}, it can also be seen that as the size of the graph increases, 
the performance of the learned policy (black) increasingly converges to the mean field system performance (red), 
validating the accuracy of our mean field formulation and choice of $N$ for training.

\begin{figure}
\centering

\includegraphics[width=\linewidth]{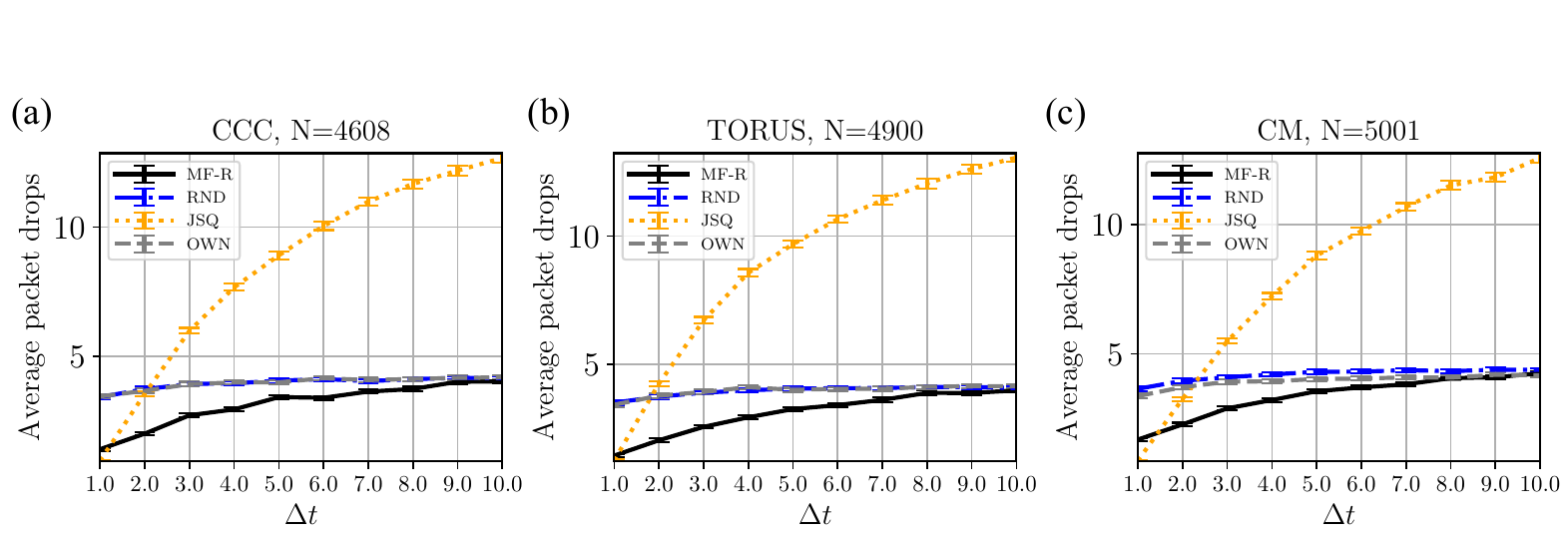}
\caption{Performance over $\Delta t\in\{1,2,\ldots, 10\}$ for large sparse graphs with underlying topologies of CCC in (a), TORUS in (b) and CM in (c). The number of nodes used to generate the graphs is given at the top of each subfigure. 
}
\label{fig:all_delta_t}
\end{figure}

\paragraph{Performance in large systems}
In Figure~\ref{fig:all_delta_t} each subfigure shows results for a different topology. We show the performance over a range of $\Delta t \in \{1,2,\ldots,10\}$ for large sparse networks. 
Figure~\ref{fig:all_delta_t}(a) is for the CCC graph with $d=3$ and $N=4608$. Figure~\ref{fig:all_delta_t}(b) shows the performance for the TORUS graph with $d=4$, $o_t=70$ and $N=4900$, while Figure~\ref{fig:all_delta_t}(c) is for the CM graph for $N=5001$ with uniformly distributed degrees in the set $d\in\{2,3\}$. Our learned MF-R policy is compared to JSQ, RND and OWN algorithms. 
The following observations can be made for all the subfigures: (i) For small $\Delta t=1$ JSQ performs better than all the algorithms which is theoretically guaranteed, however performance of MF-R is very close to it; (ii) For in between range of $\Delta t = \{2,3,\ldots,7\}$ MF-R has the best performance in terms of packets dropped; (iii) Except in the CM graph (where the varying degrees lead to differing packet arrival rates between OWN and RND), we find that OWN and RND again have equivalent performance by regularity of the graph, as discussed in the previous paragraph; (iv) For higher $\Delta t \in\{8,9,10\}$ our MF-R policy performance coincides with the RND policy, where the RND policy has been theoretically proven to be optimal as the synchronization delay $\Delta t \to \infty$ \cite{mitzenmacher2001power}. Hence, we find that MF-R policy performs consistently well in various topologies.




\begin{figure}
\centering
\includegraphics[width=0.5\linewidth]{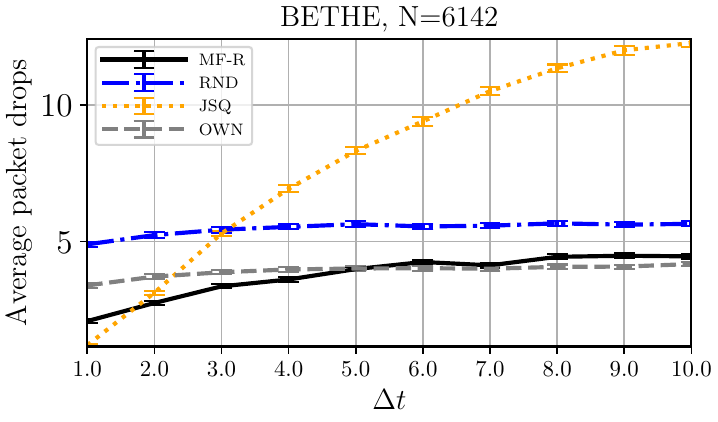}
\caption{Performance comparison on large-sized Bethe lattice graph. The MF-R can be worse than RND due to violation of modelling assumptions.}
\label{fig:bethe}
\end{figure}

\paragraph{Ablation on topological assumptions}
As mentioned in Section\ref{sec:topologies}, the theoretical analysis and work in \cite{lacker2019local} does not apply to the Bethe topology, and the effect is further supported by our experimental results.
Figure \ref{fig:bethe} shows the performance comparison for Bethe lattice graphs of the learned MF-R policy with JSQ, RND, and OWN algorithms over different $\Delta t$. We show the result only for a large Bethe network, with $o_b=11$, $d=3$ and $N=6142$. OWN outperforms RND more significantly than in CM, since disproportionately many packets arrive at nodes connected to leaf nodes.
It can be seen that the system behaves as expected when $\Delta t$ is small, with JSQ quickly outperformed by MF-R, while MF-R eventually performs worse than OWN as $\Delta t$ increases. This behaviour is due to the fact that in structures such as regular trees, the central node as the root node is not sufficient to represent the performance of all nodes, especially the many leaf nodes that behave differently even in the limit. Because of this loss of regularity, theoretical guarantees fail when scaling large Bethe or similar topologies, as explained in Section \ref{sec:topologies}.
Even so, we see that the learned policy can outperform existing solutions in certain regimes, e.g., at $\Delta t = 3$.

For the sake of completeness we have also provided the results of following experiments in Appendix: (i) convergence analysis of our simulator, (ii) effect of different neural network parameters, (iii) two other MF based policies (iv) comparison to the SAC algorithm \cite{yiheng2021}, (v) effect of increasing buffer size, (vi) having heterogenous servers, (vi) using recurrent neural networks with PPO and (vii) use of different observation spaces for MF policies.

\section{Conclusion}
Overall, we have learned efficient algorithms for load balancing in highly local queuing systems with some degree of theoretical guarantees through sparse mean-field approximations. The approach has been positively evaluated in comparison to well-known baselines, and can scale to queueing systems of arbitrary size with arbitrary wireless communication delays.
Future work could attempt to quantify rates of convergence for approximate optimality guarantees of MFC policies under the framework of \cite{lacker2019local}, or consider dynamic programming principles.
One could also look into the analysis of graph structures such as Bethe and regular trees, for which the current 
sparse MFC modelling does not suffice. 



\appendix

\bibliographystyle{ACM-Reference-Format}
\bibliography{references}


\begin{thebibliography}{44}


\ifx \showCODEN    \undefined \def \showCODEN     #1{\unskip}     \fi
\ifx \showDOI      \undefined \def \showDOI       #1{#1}\fi
\ifx \showISBNx    \undefined \def \showISBNx     #1{\unskip}     \fi
\ifx \showISBNxiii \undefined \def \showISBNxiii  #1{\unskip}     \fi
\ifx \showISSN     \undefined \def \showISSN      #1{\unskip}     \fi
\ifx \showLCCN     \undefined \def \showLCCN      #1{\unskip}     \fi
\ifx \shownote     \undefined \def \shownote      #1{#1}          \fi
\ifx \showarticletitle \undefined \def \showarticletitle #1{#1}   \fi
\ifx \showURL      \undefined \def \showURL       {\relax}        \fi
\providecommand\bibfield[2]{#2}
\providecommand\bibinfo[2]{#2}
\providecommand\natexlab[1]{#1}
\providecommand\showeprint[2][]{arXiv:#2}

\bibitem[\protect\citeauthoryear{Bensoussan, Frehse, Yam, et~al\mbox{.}}{Bensoussan et~al\mbox{.}}{2013}]%
        {bensoussan2013mean}
\bibfield{author}{\bibinfo{person}{Alain Bensoussan}, \bibinfo{person}{Jens Frehse}, \bibinfo{person}{Phillip Yam}, {et~al\mbox{.}}} \bibinfo{year}{2013}\natexlab{}.
\newblock \bibinfo{booktitle}{{\em Mean field games and mean field type control theory}}. Vol.~\bibinfo{volume}{101}.
\newblock \bibinfo{publisher}{Springer}.
\newblock


\bibitem[\protect\citeauthoryear{Caines and Huang}{Caines and Huang}{2021}]%
        {caines2021graphon}
\bibfield{author}{\bibinfo{person}{Peter~E Caines} {and} \bibinfo{person}{Minyi Huang}.} \bibinfo{year}{2021}\natexlab{}.
\newblock \showarticletitle{Graphon mean field games and their equations}.
\newblock \bibinfo{journal}{{\em SIAM Journal on Control and Optimization\/}} \bibinfo{volume}{59}, \bibinfo{number}{6} (\bibinfo{year}{2021}), \bibinfo{pages}{4373--4399}.
\newblock


\bibitem[\protect\citeauthoryear{Christianos, Sch{\"a}fer, and Albrecht}{Christianos et~al\mbox{.}}{2020}]%
        {christianos2020shared}
\bibfield{author}{\bibinfo{person}{Filippos Christianos}, \bibinfo{person}{Lukas Sch{\"a}fer}, {and} \bibinfo{person}{Stefano Albrecht}.} \bibinfo{year}{2020}\natexlab{}.
\newblock \showarticletitle{Shared experience actor-critic for multi-agent reinforcement learning}.
\newblock \bibinfo{journal}{{\em Advances in Neural Information Processing Systems\/}}  \bibinfo{volume}{33} (\bibinfo{year}{2020}), \bibinfo{pages}{10707--10717}.
\newblock


\bibitem[\protect\citeauthoryear{Chu, Chinchali, and Katti}{Chu et~al\mbox{.}}{2020}]%
        {chu2020multi}
\bibfield{author}{\bibinfo{person}{Tianshu Chu}, \bibinfo{person}{Sandeep Chinchali}, {and} \bibinfo{person}{Sachin Katti}.} \bibinfo{year}{2020}\natexlab{}.
\newblock \showarticletitle{Multi-agent Reinforcement Learning for Networked System Control}. In \bibinfo{booktitle}{{\em 8th International Conference on Learning Representations}}. \bibinfo{publisher}{OpenReview.net}, \bibinfo{address}{Addis Ababa, Ethiopia, April 26-30, 2020}, \bibinfo{pages}{1--17}.
\newblock


\bibitem[\protect\citeauthoryear{Cui and Koeppl}{Cui and Koeppl}{2022}]%
        {cui2021learning}
\bibfield{author}{\bibinfo{person}{Kai Cui} {and} \bibinfo{person}{Heinz Koeppl}.} \bibinfo{year}{2022}\natexlab{}.
\newblock \showarticletitle{Learning Graphon Mean Field Games and Approximate Nash Equilibria}. In \bibinfo{booktitle}{{\em The 10th International Conference on Learning Representations}}. \bibinfo{publisher}{OpenReview.net}, \bibinfo{pages}{1--31}.
\newblock


\bibitem[\protect\citeauthoryear{Cui, Tahir, Ekinci, Elshamanhory, Eich, Li, and Koeppl}{Cui et~al\mbox{.}}{2022}]%
        {cui2022survey}
\bibfield{author}{\bibinfo{person}{Kai Cui}, \bibinfo{person}{Anam Tahir}, \bibinfo{person}{Gizem Ekinci}, \bibinfo{person}{Ahmed Elshamanhory}, \bibinfo{person}{Yannick Eich}, \bibinfo{person}{Mengguang Li}, {and} \bibinfo{person}{Heinz Koeppl}.} \bibinfo{year}{2022}\natexlab{}.
\newblock \showarticletitle{A Survey on Large-Population Systems and Scalable Multi-Agent Reinforcement Learning}.
\newblock \bibinfo{journal}{{\em arXiv preprint arXiv:2209.03859\/}} (\bibinfo{year}{2022}), \bibinfo{pages}{1--21}.
\newblock


\bibitem[\protect\citeauthoryear{Dawson, Tang, and Zhao}{Dawson et~al\mbox{.}}{2005}]%
        {dawson2005balancing}
\bibfield{author}{\bibinfo{person}{Donald~A Dawson}, \bibinfo{person}{Jiashan Tang}, {and} \bibinfo{person}{Yiqiang~Q Zhao}.} \bibinfo{year}{2005}\natexlab{}.
\newblock \showarticletitle{Balancing queues by mean field interaction}.
\newblock \bibinfo{journal}{{\em Queueing Systems\/}} \bibinfo{volume}{49}, \bibinfo{number}{3} (\bibinfo{year}{2005}), \bibinfo{pages}{335--361}.
\newblock


\bibitem[\protect\citeauthoryear{der Boor, Borst, Van~Leeuwaarden, and Mukherjee}{der Boor et~al\mbox{.}}{2022}]%
        {der2022scalable}
\bibfield{author}{\bibinfo{person}{Mark~Van der Boor}, \bibinfo{person}{Sem~C Borst}, \bibinfo{person}{Johan~SH Van~Leeuwaarden}, {and} \bibinfo{person}{Debankur Mukherjee}.} \bibinfo{year}{2022}\natexlab{}.
\newblock \showarticletitle{Scalable load balancing in networked systems: A survey of recent advances}.
\newblock \bibinfo{journal}{{\it SIAM Rev.}} \bibinfo{volume}{64}, \bibinfo{number}{3} (\bibinfo{year}{2022}), \bibinfo{pages}{554--622}.
\newblock


\bibitem[\protect\citeauthoryear{Deveci, Devine, Pedretti, Taylor, Rajamanickam, and {\c{C}}ataly{\"u}rek}{Deveci et~al\mbox{.}}{2019}]%
        {mehmet2019networks}
\bibfield{author}{\bibinfo{person}{Mehmet Deveci}, \bibinfo{person}{Karen~D Devine}, \bibinfo{person}{Kevin Pedretti}, \bibinfo{person}{Mark~A Taylor}, \bibinfo{person}{Sivasankaran Rajamanickam}, {and} \bibinfo{person}{{\"U}mit~V {\c{C}}ataly{\"u}rek}.} \bibinfo{year}{2019}\natexlab{}.
\newblock \showarticletitle{Geometric mapping of tasks to processors on parallel computers with mesh or torus networks}.
\newblock \bibinfo{journal}{{\em IEEE Transactions on Parallel and Distributed Systems\/}} \bibinfo{volume}{30}, \bibinfo{number}{9} (\bibinfo{year}{2019}), \bibinfo{pages}{2018--2032}.
\newblock


\bibitem[\protect\citeauthoryear{Fabian, Cui, and Koeppl}{Fabian et~al\mbox{.}}{2022}]%
        {fabian2022learning}
\bibfield{author}{\bibinfo{person}{Christian Fabian}, \bibinfo{person}{Kai Cui}, {and} \bibinfo{person}{Heinz Koeppl}.} \bibinfo{year}{2022}\natexlab{}.
\newblock \showarticletitle{Learning Sparse Graphon Mean Field Games}.
\newblock \bibinfo{journal}{{\em arXiv preprint arXiv:2209.03880\/}} (\bibinfo{year}{2022}), \bibinfo{pages}{1--32}.
\newblock


\bibitem[\protect\citeauthoryear{Fischer and Meier-Hellstern}{Fischer and Meier-Hellstern}{1993}]%
        {fischer1993markov}
\bibfield{author}{\bibinfo{person}{Wolfgang Fischer} {and} \bibinfo{person}{Kathleen Meier-Hellstern}.} \bibinfo{year}{1993}\natexlab{}.
\newblock \showarticletitle{The Markov-modulated Poisson process (MMPP) cookbook}.
\newblock \bibinfo{journal}{{\em Performance evaluation\/}} \bibinfo{volume}{18}, \bibinfo{number}{2} (\bibinfo{year}{1993}), \bibinfo{pages}{149--171}.
\newblock


\bibitem[\protect\citeauthoryear{Fosdick, Larremore, Nishimura, and Ugander}{Fosdick et~al\mbox{.}}{2018}]%
        {fosdick2018configuring}
\bibfield{author}{\bibinfo{person}{Bailey~K Fosdick}, \bibinfo{person}{Daniel~B Larremore}, \bibinfo{person}{Joel Nishimura}, {and} \bibinfo{person}{Johan Ugander}.} \bibinfo{year}{2018}\natexlab{}.
\newblock \showarticletitle{Configuring random graph models with fixed degree sequences}.
\newblock \bibinfo{journal}{{\it SIAM Rev.}} \bibinfo{volume}{60}, \bibinfo{number}{2} (\bibinfo{year}{2018}), \bibinfo{pages}{315--355}.
\newblock


\bibitem[\protect\citeauthoryear{Gast}{Gast}{2015}]%
        {gast2015power}
\bibfield{author}{\bibinfo{person}{Nicolas Gast}.} \bibinfo{year}{2015}\natexlab{}.
\newblock \showarticletitle{The power of two choices on graphs: The pair-approximation is accurate?}
\newblock \bibinfo{journal}{{\em ACM SIGMETRICS Performance Evaluation Review\/}} \bibinfo{volume}{43}, \bibinfo{number}{2} (\bibinfo{year}{2015}), \bibinfo{pages}{69--71}.
\newblock


\bibitem[\protect\citeauthoryear{Guicheng and Yang}{Guicheng and Yang}{2022}]%
        {guicheng2022review}
\bibfield{author}{\bibinfo{person}{Shen Guicheng} {and} \bibinfo{person}{Wang Yang}.} \bibinfo{year}{2022}\natexlab{}.
\newblock \showarticletitle{Review on {Dec-POMDP} Model for {MARL} Algorithms}.
\newblock In \bibinfo{booktitle}{{\em Smart Communications, Intelligent Algorithms and Interactive Methods}}. \bibinfo{publisher}{Springer}, \bibinfo{pages}{29--35}.
\newblock


\bibitem[\protect\citeauthoryear{Habibian and Patooghy}{Habibian and Patooghy}{2017}]%
        {habibian2017fault}
\bibfield{author}{\bibinfo{person}{Hossein Habibian} {and} \bibinfo{person}{Ahmad Patooghy}.} \bibinfo{year}{2017}\natexlab{}.
\newblock \showarticletitle{Fault-tolerant routing methodology for hypercube and cube-connected cycles interconnection networks}.
\newblock \bibinfo{journal}{{\em The Journal of Supercomputing\/}} \bibinfo{volume}{73}, \bibinfo{number}{10} (\bibinfo{year}{2017}), \bibinfo{pages}{4560--4579}.
\newblock


\bibitem[\protect\citeauthoryear{Harremo{\"e}s, Johnson, and Kontoyiannis}{Harremo{\"e}s et~al\mbox{.}}{2007}]%
        {harremoes2007thinning}
\bibfield{author}{\bibinfo{person}{Peter Harremo{\"e}s}, \bibinfo{person}{Oliver Johnson}, {and} \bibinfo{person}{Ioannis Kontoyiannis}.} \bibinfo{year}{2007}\natexlab{}.
\newblock \showarticletitle{Thinning and the law of small numbers}. In \bibinfo{booktitle}{{\em IEEE International Symposium on Information Theory}}. \bibinfo{publisher}{IEEE}, \bibinfo{pages}{1491--1495}.
\newblock


\bibitem[\protect\citeauthoryear{Hu, Wei, Yan, and Zhang}{Hu et~al\mbox{.}}{2022}]%
        {hu2022graphon}
\bibfield{author}{\bibinfo{person}{Yuanquan Hu}, \bibinfo{person}{Xiaoli Wei}, \bibinfo{person}{Junji Yan}, {and} \bibinfo{person}{Hengxi Zhang}.} \bibinfo{year}{2022}\natexlab{}.
\newblock \showarticletitle{Graphon Mean-Field Control for Cooperative Multi-Agent Reinforcement Learning}.
\newblock \bibinfo{journal}{{\em arXiv preprint arXiv:2209.04808\/}} (\bibinfo{year}{2022}), \bibinfo{pages}{1--25}.
\newblock


\bibitem[\protect\citeauthoryear{Lacker, Ramanan, and Wu}{Lacker et~al\mbox{.}}{2023}]%
        {lacker2019local}
\bibfield{author}{\bibinfo{person}{Daniel Lacker}, \bibinfo{person}{Kavita Ramanan}, {and} \bibinfo{person}{Ruoyu Wu}.} \bibinfo{year}{2023}\natexlab{}.
\newblock \showarticletitle{Local weak convergence for sparse networks of interacting processes}.
\newblock \bibinfo{journal}{{\em The Annals of Applied Probability\/}} \bibinfo{volume}{33}, \bibinfo{number}{2} (\bibinfo{year}{2023}), \bibinfo{pages}{843--888}.
\newblock


\bibitem[\protect\citeauthoryear{Lasry and Lions}{Lasry and Lions}{2007}]%
        {lasry2007mean}
\bibfield{author}{\bibinfo{person}{Jean-Michel Lasry} {and} \bibinfo{person}{Pierre-Louis Lions}.} \bibinfo{year}{2007}\natexlab{}.
\newblock \showarticletitle{Mean field games}.
\newblock \bibinfo{journal}{{\em Japanese Journal of Mathematics\/}} \bibinfo{volume}{2}, \bibinfo{number}{1} (\bibinfo{year}{2007}), \bibinfo{pages}{229--260}.
\newblock


\bibitem[\protect\citeauthoryear{Li}{Li}{2021}]%
        {yiheng2021}
\bibfield{author}{\bibinfo{person}{Yiheng Li}.} \bibinfo{year}{2021}\natexlab{}.
\newblock \bibinfo{title}{Networked-MARL}.
\newblock \bibinfo{howpublished}{\url{https://github.com/yihenglin97/Networked-MARL}}.   (\bibinfo{year}{2021}).
\newblock


\bibitem[\protect\citeauthoryear{Liang, Liaw, Nishihara, Moritz, Fox, Goldberg, Gonzalez, Jordan, and Stoica}{Liang et~al\mbox{.}}{2018}]%
        {liang2018rllib}
\bibfield{author}{\bibinfo{person}{Eric Liang}, \bibinfo{person}{Richard Liaw}, \bibinfo{person}{Robert Nishihara}, \bibinfo{person}{Philipp Moritz}, \bibinfo{person}{Roy Fox}, \bibinfo{person}{Ken Goldberg}, \bibinfo{person}{Joseph Gonzalez}, \bibinfo{person}{Michael Jordan}, {and} \bibinfo{person}{Ion Stoica}.} \bibinfo{year}{2018}\natexlab{}.
\newblock \showarticletitle{RLlib: Abstractions for distributed reinforcement learning}. In \bibinfo{booktitle}{{\em International Conference on Machine Learning}}. \bibinfo{publisher}{PMLR}, \bibinfo{pages}{3053--3062}.
\newblock


\bibitem[\protect\citeauthoryear{Lin, Qu, Huang, and Wierman}{Lin et~al\mbox{.}}{2021}]%
        {lin2021multi}
\bibfield{author}{\bibinfo{person}{Yiheng Lin}, \bibinfo{person}{Guannan Qu}, \bibinfo{person}{Longbo Huang}, {and} \bibinfo{person}{Adam Wierman}.} \bibinfo{year}{2021}\natexlab{}.
\newblock \showarticletitle{Multi-agent reinforcement learning in stochastic networked systems}.
\newblock \bibinfo{journal}{{\em Advances in Neural Information Processing Systems\/}}  \bibinfo{volume}{34} (\bibinfo{year}{2021}), \bibinfo{pages}{7825--7837}.
\newblock


\bibitem[\protect\citeauthoryear{Lipshutz}{Lipshutz}{2019}]%
        {lipshutz2019open}
\bibfield{author}{\bibinfo{person}{David Lipshutz}.} \bibinfo{year}{2019}\natexlab{}.
\newblock \showarticletitle{Open problem—load balancing using delayed information}.
\newblock \bibinfo{journal}{{\em Stochastic Systems\/}} \bibinfo{volume}{9}, \bibinfo{number}{3} (\bibinfo{year}{2019}), \bibinfo{pages}{305--306}.
\newblock


\bibitem[\protect\citeauthoryear{Lov{\'a}sz}{Lov{\'a}sz}{2012}]%
        {lovasz2012large}
\bibfield{author}{\bibinfo{person}{L{\'a}szl{\'o} Lov{\'a}sz}.} \bibinfo{year}{2012}\natexlab{}.
\newblock \bibinfo{booktitle}{{\em Large networks and graph limits}}. Vol.~\bibinfo{volume}{60}.
\newblock \bibinfo{publisher}{American Mathematical Society}.
\newblock


\bibitem[\protect\citeauthoryear{Lu, Xie, Kliot, Geller, Larus, and Greenberg}{Lu et~al\mbox{.}}{2011}]%
        {lu2011join}
\bibfield{author}{\bibinfo{person}{Yi Lu}, \bibinfo{person}{Qiaomin Xie}, \bibinfo{person}{Gabriel Kliot}, \bibinfo{person}{Alan Geller}, \bibinfo{person}{James~R Larus}, {and} \bibinfo{person}{Albert Greenberg}.} \bibinfo{year}{2011}\natexlab{}.
\newblock \showarticletitle{Join-Idle-Queue: A novel load balancing algorithm for dynamically scalable web services}.
\newblock \bibinfo{journal}{{\em Performance Evaluation\/}} \bibinfo{volume}{68}, \bibinfo{number}{11} (\bibinfo{year}{2011}), \bibinfo{pages}{1056--1071}.
\newblock


\bibitem[\protect\citeauthoryear{Mishra, Sahoo, and Parida}{Mishra et~al\mbox{.}}{2020}]%
        {mishra2020load}
\bibfield{author}{\bibinfo{person}{Sambit~Kumar Mishra}, \bibinfo{person}{Bibhudatta Sahoo}, {and} \bibinfo{person}{Priti~Paramita Parida}.} \bibinfo{year}{2020}\natexlab{}.
\newblock \showarticletitle{Load balancing in cloud computing: a big picture}.
\newblock \bibinfo{journal}{{\em Journal of King Saud University-Computer and Information Sciences\/}} \bibinfo{volume}{32}, \bibinfo{number}{2} (\bibinfo{year}{2020}), \bibinfo{pages}{149--158}.
\newblock


\bibitem[\protect\citeauthoryear{Mitzenmacher}{Mitzenmacher}{2001}]%
        {mitzenmacher2001power}
\bibfield{author}{\bibinfo{person}{Michael Mitzenmacher}.} \bibinfo{year}{2001}\natexlab{}.
\newblock \showarticletitle{The power of two choices in randomized load balancing}.
\newblock \bibinfo{journal}{{\em IEEE Transactions on Parallel and Distributed Systems\/}} \bibinfo{volume}{12}, \bibinfo{number}{10} (\bibinfo{year}{2001}), \bibinfo{pages}{1094--1104}.
\newblock


\bibitem[\protect\citeauthoryear{Mukherjee, Borst, and Van~Leeuwaarden}{Mukherjee et~al\mbox{.}}{2018}]%
        {mukherjee2018asymptotically}
\bibfield{author}{\bibinfo{person}{Debankur Mukherjee}, \bibinfo{person}{Sem~C Borst}, {and} \bibinfo{person}{Johan~SH Van~Leeuwaarden}.} \bibinfo{year}{2018}\natexlab{}.
\newblock \showarticletitle{Asymptotically optimal load balancing topologies}.
\newblock \bibinfo{journal}{{\em Proceedings of the ACM on Measurement and Analysis of Computing Systems\/}} \bibinfo{volume}{2}, \bibinfo{number}{1} (\bibinfo{year}{2018}), \bibinfo{pages}{1--29}.
\newblock


\bibitem[\protect\citeauthoryear{Ni, Eysenbach, and Salakhutdinov}{Ni et~al\mbox{.}}{2022}]%
        {ni2021recurrent}
\bibfield{author}{\bibinfo{person}{Tianwei Ni}, \bibinfo{person}{Benjamin Eysenbach}, {and} \bibinfo{person}{Ruslan Salakhutdinov}.} \bibinfo{year}{2022}\natexlab{}.
\newblock \showarticletitle{Recurrent Model-Free {RL} Can Be a Strong Baseline for Many {POMDP}s}. In \bibinfo{booktitle}{{\em Proceedings of the 39th International Conference on Machine Learning}}, \bibfield{editor}{\bibinfo{person}{Kamalika Chaudhuri}, \bibinfo{person}{Stefanie Jegelka}, \bibinfo{person}{Le~Song}, \bibinfo{person}{Csaba Szepesvari}, \bibinfo{person}{Gang Niu}, {and} \bibinfo{person}{Sivan Sabato}} (Eds.), Vol.~\bibinfo{volume}{162}. \bibinfo{publisher}{PMLR}, \bibinfo{pages}{16691--16723}.
\newblock


\bibitem[\protect\citeauthoryear{Papoudakis, Christianos, Sch{\"a}fer, and Albrecht}{Papoudakis et~al\mbox{.}}{2021}]%
        {papoudakis2021benchmarking}
\bibfield{author}{\bibinfo{person}{Georgios Papoudakis}, \bibinfo{person}{Filippos Christianos}, \bibinfo{person}{Lukas Sch{\"a}fer}, {and} \bibinfo{person}{Stefano~V Albrecht}.} \bibinfo{year}{2021}\natexlab{}.
\newblock \showarticletitle{Benchmarking multi-agent deep reinforcement learning algorithms in cooperative tasks}. In \bibinfo{booktitle}{{\em Proceedings NeurIPS Datasets and Benchmarks}}. \bibinfo{publisher}{The MIT Press}, \bibinfo{pages}{15--19}.
\newblock


\bibitem[\protect\citeauthoryear{Pham and Wei}{Pham and Wei}{2018}]%
        {pham2018bellman}
\bibfield{author}{\bibinfo{person}{Huy{\^e}n Pham} {and} \bibinfo{person}{Xiaoli Wei}.} \bibinfo{year}{2018}\natexlab{}.
\newblock \showarticletitle{Bellman equation and viscosity solutions for mean-field stochastic control problem}.
\newblock \bibinfo{journal}{{\em ESAIM: Control, Optimisation and Calculus of Variations\/}} \bibinfo{volume}{24}, \bibinfo{number}{1} (\bibinfo{year}{2018}), \bibinfo{pages}{437--461}.
\newblock


\bibitem[\protect\citeauthoryear{Rodriguez and Guillemin}{Rodriguez and Guillemin}{2018}]%
        {rodriguez2018cloud}
\bibfield{author}{\bibinfo{person}{Veronica~Quintuna Rodriguez} {and} \bibinfo{person}{Fabrice Guillemin}.} \bibinfo{year}{2018}\natexlab{}.
\newblock \showarticletitle{Cloud-RAN modeling based on parallel processing}.
\newblock \bibinfo{journal}{{\em IEEE Journal on Selected Areas in Communications\/}} \bibinfo{volume}{36}, \bibinfo{number}{3} (\bibinfo{year}{2018}), \bibinfo{pages}{457--468}.
\newblock


\bibitem[\protect\citeauthoryear{Rutten and Mukherjee}{Rutten and Mukherjee}{2023}]%
        {rutten2023mean}
\bibfield{author}{\bibinfo{person}{Daan Rutten} {and} \bibinfo{person}{Debankur Mukherjee}.} \bibinfo{year}{2023}\natexlab{}.
\newblock \showarticletitle{Mean-field analysis for load balancing on spatial graphs}.
\newblock \bibinfo{journal}{{\em arXiv preprint arXiv:2301.03493\/}} (\bibinfo{year}{2023}), \bibinfo{pages}{1--27}.
\newblock


\bibitem[\protect\citeauthoryear{Schulman, Wolski, Dhariwal, Radford, and Klimov}{Schulman et~al\mbox{.}}{2017}]%
        {schulman2017proximal}
\bibfield{author}{\bibinfo{person}{John Schulman}, \bibinfo{person}{Filip Wolski}, \bibinfo{person}{Prafulla Dhariwal}, \bibinfo{person}{Alec Radford}, {and} \bibinfo{person}{Oleg Klimov}.} \bibinfo{year}{2017}\natexlab{}.
\newblock \showarticletitle{Proximal policy optimization algorithms}.
\newblock \bibinfo{journal}{{\em arXiv preprint arXiv:1707.06347\/}} (\bibinfo{year}{2017}), \bibinfo{pages}{1--12}.
\newblock


\bibitem[\protect\citeauthoryear{Selen, Adan, Kapodistria, and van Leeuwaarden}{Selen et~al\mbox{.}}{2016}]%
        {selen2016steady}
\bibfield{author}{\bibinfo{person}{Jori Selen}, \bibinfo{person}{Ivo Adan}, \bibinfo{person}{Stella Kapodistria}, {and} \bibinfo{person}{Johan van Leeuwaarden}.} \bibinfo{year}{2016}\natexlab{}.
\newblock \showarticletitle{Steady-state analysis of shortest expected delay routing}.
\newblock \bibinfo{journal}{{\em Queueing Systems\/}} \bibinfo{volume}{84}, \bibinfo{number}{3} (\bibinfo{year}{2016}), \bibinfo{pages}{309--354}.
\newblock


\bibitem[\protect\citeauthoryear{Shortle, Thompson, Gross, and Harris}{Shortle et~al\mbox{.}}{2018}]%
        {shortle2018fundamentals}
\bibfield{author}{\bibinfo{person}{John~F Shortle}, \bibinfo{person}{James~M Thompson}, \bibinfo{person}{Donald Gross}, {and} \bibinfo{person}{Carl~M Harris}.} \bibinfo{year}{2018}\natexlab{}.
\newblock \bibinfo{booktitle}{{\em Fundamentals of queueing theory}}. Vol.~\bibinfo{volume}{399}.
\newblock \bibinfo{publisher}{John Wiley \& Sons}.
\newblock


\bibitem[\protect\citeauthoryear{Sutton and Barto}{Sutton and Barto}{2018}]%
        {sutton2018reinforcement}
\bibfield{author}{\bibinfo{person}{Richard~S Sutton} {and} \bibinfo{person}{Andrew~G Barto}.} \bibinfo{year}{2018}\natexlab{}.
\newblock \bibinfo{booktitle}{{\em Reinforcement learning: An introduction}}.
\newblock \bibinfo{publisher}{MIT press}.
\newblock


\bibitem[\protect\citeauthoryear{Szab{\'o} and Borsos}{Szab{\'o} and Borsos}{2016}]%
        {szabo2016evolutionary}
\bibfield{author}{\bibinfo{person}{Gy{\"o}rgy Szab{\'o}} {and} \bibinfo{person}{Istv{\'a}n Borsos}.} \bibinfo{year}{2016}\natexlab{}.
\newblock \showarticletitle{Evolutionary potential games on lattices}.
\newblock \bibinfo{journal}{{\em Physics Reports\/}}  \bibinfo{volume}{624} (\bibinfo{year}{2016}), \bibinfo{pages}{1--60}.
\newblock


\bibitem[\protect\citeauthoryear{Tahir, Cui, and Koeppl}{Tahir et~al\mbox{.}}{2022}]%
        {tahir2022learning}
\bibfield{author}{\bibinfo{person}{Anam Tahir}, \bibinfo{person}{Kai Cui}, {and} \bibinfo{person}{Heinz Koeppl}.} \bibinfo{year}{2022}\natexlab{}.
\newblock \showarticletitle{Learning mean-field control for delayed information load balancing in large queuing systems}. In \bibinfo{booktitle}{{\em Proceedings of the 51st International Conference on Parallel Processing}}. \bibinfo{publisher}{ACM New York, NY, USA}, \bibinfo{pages}{1--11}.
\newblock


\bibitem[\protect\citeauthoryear{Walker, Bora, and Fidler}{Walker et~al\mbox{.}}{2022}]%
        {walker2022performance}
\bibfield{author}{\bibinfo{person}{Brenton Walker}, \bibinfo{person}{Stefan Bora}, {and} \bibinfo{person}{Markus Fidler}.} \bibinfo{year}{2022}\natexlab{}.
\newblock \showarticletitle{Performance and scaling of parallel systems with blocking start and/or departure barriers}. In \bibinfo{booktitle}{{\em IEEE Conference on Computer Communications}}. \bibinfo{publisher}{IEEE}, \bibinfo{pages}{460--469}.
\newblock


\bibitem[\protect\citeauthoryear{Winston}{Winston}{1977}]%
        {winston1977optimality}
\bibfield{author}{\bibinfo{person}{Wayne Winston}.} \bibinfo{year}{1977}\natexlab{}.
\newblock \showarticletitle{Optimality of the shortest line discipline}.
\newblock \bibinfo{journal}{{\em Journal of Applied Probability\/}} \bibinfo{volume}{14}, \bibinfo{number}{1} (\bibinfo{year}{1977}), \bibinfo{pages}{181--189}.
\newblock


\bibitem[\protect\citeauthoryear{Yu, Velu, Vinitsky, Wang, Bayen, and Wu}{Yu et~al\mbox{.}}{2022}]%
        {yu2021surprising}
\bibfield{author}{\bibinfo{person}{Chao Yu}, \bibinfo{person}{Akash Velu}, \bibinfo{person}{Eugene Vinitsky}, \bibinfo{person}{Yu Wang}, \bibinfo{person}{Alexandre Bayen}, {and} \bibinfo{person}{Yi Wu}.} \bibinfo{year}{2022}\natexlab{}.
\newblock \showarticletitle{The surprising effectiveness of {PPO} in cooperative, multi-agent games}.
\newblock \bibinfo{journal}{{\em Proceedings NeurIPS Datasets and Benchmarks\/}} (\bibinfo{year}{2022}), \bibinfo{pages}{1--14}.
\newblock


\bibitem[\protect\citeauthoryear{Zhang, Yang, and Ba{\c{s}}ar}{Zhang et~al\mbox{.}}{2021}]%
        {zhang2021multi}
\bibfield{author}{\bibinfo{person}{Kaiqing Zhang}, \bibinfo{person}{Zhuoran Yang}, {and} \bibinfo{person}{Tamer Ba{\c{s}}ar}.} \bibinfo{year}{2021}\natexlab{}.
\newblock \showarticletitle{Multi-agent reinforcement learning: A selective overview of theories and algorithms}.
\newblock \bibinfo{journal}{{\em Handbook of Reinforcement Learning and Control\/}} (\bibinfo{year}{2021}), \bibinfo{pages}{321--384}.
\newblock


\bibitem[\protect\citeauthoryear{Zhou, Shroff, and Wierman}{Zhou et~al\mbox{.}}{2021}]%
        {zhou2021asymptotically}
\bibfield{author}{\bibinfo{person}{Xingyu Zhou}, \bibinfo{person}{Ness Shroff}, {and} \bibinfo{person}{Adam Wierman}.} \bibinfo{year}{2021}\natexlab{}.
\newblock \showarticletitle{Asymptotically optimal load balancing in large-scale heterogeneous systems with multiple dispatchers}.
\newblock \bibinfo{journal}{{\em Performance Evaluation\/}}  \bibinfo{volume}{145} (\bibinfo{year}{2021}), \bibinfo{pages}{102146}.
\newblock


\end{thebibliography}

\newpage
\appendix









\section{Appendix}
This Appendix contains the proofs, algorithm, hyperparameter values and additional experiments connected to the paper titled: "Sparse Mean Field Load Balancing in Large Localized Queueing Systems".
\section{Proof}
\begin{proof}[Proof of Theorem 2.1]
We apply the framework of \cite{lacker2019local}, which does not include actions or rewards, by including actions and rewards via separate time steps. Specifically, each time step is split in three, and we define the agent state space $\mathcal X \coloneqq \mathcal Z \cup (\mathcal Z \times \mathcal A) \cup (\mathcal Z \times \mathcal A \times [0, D_{\mathrm{max}}])$ for each of the agents, indicating at any third decision epoch $t=0, 3, 6, \ldots$ the state of its queue. After each such epoch, the agent states will contain the state of its queue together with its choice of action at times $t=1, 4, \ldots$, and lastly also the number of packet drops at times $t=2,5,\ldots$. Here, $D_{\mathrm{max}}$ is the maximum expected number of packet drops and is given by the $(d+1)$ times the maximum per-scheduler arrival rate $\lambda_{\mathrm{max}} = \lambda_h$. 

We formally rewrite the system defined earlier, using the symbol $\mathbf x$ for the state of the rewritten system instead of $\mathbf z$. As a result, each agent $i$ is endowed with a random local $\mathcal X$-valued state variable $X^{G,x}_i(t)$ at each time $t$. For the dynamics, we let $S^{\sqcup}(\mathcal X)$ denote the space of unordered terminating sequences (up to some maximum degree) with the discrete topology as in \cite{lacker2019local}, and define in the following a system dynamics function $F \colon \mathcal X \times S^{\sqcup}(\mathcal X) \times \Xi \to \mathcal X$ returning a new state for any current local state $X^{G,x}_i(t)$, any current $S^{\sqcup}(\mathcal X)$-valued neighbourhood queue fillings $X^{G,x}_{N_i}(t)$ and i.i.d. sampled $\Xi$-valued noise $\xi_{i}(t+1)$. More precisely, we use
\begin{align*}
    &X^{G,x}_i(t+1) = F[X^{G,x}_i(t), X^{G,x}_{N_i}(t), \xi_{i}(t+1)] \\
    &\quad = 
    \begin{cases}
        (X^{G,x}_i(t), \xi_{i,X^{G,x}_i(t)}(t+1)), \\
        \quad \text{if} \quad X^{G,x}_i(t) \in \mathcal Z, \\
        (X^{G,x}_{i,1}(t), X^{G,x}_{i,2}(t), \xi_{i,X^{G,x}_{i,1}(t), X^{G,x}_{i,2}(t),X^{G,x}_{N_i}(t)}(t+1)) \\
        \quad \text{if} \quad X^{G,x}_i(t) \in \mathcal Z \times \mathcal A, \\
        \xi_{i,X^{G,x}_{i}(t),X^{G,x}_{N_i}(t)}(t+1), \\
        \quad \text{if} \quad X^{G,x}_i(t) \in \mathcal Z \times \mathcal A \times [0, D_{\mathrm{max}}],
    \end{cases}
\end{align*}
where we define $\Xi$ and the random noise variable $\xi$ as a contingency over all transitions, through a finite tuple of random variables $\xi_i$ consisting of components for (i) randomly sampling the next action $(\xi_{i,z})_{z \in \mathcal Z}$ according to $\pi$, (ii) computing the expected lost packets $(\xi_{i,x_1,x_2,\mu})_{(x_1,x_2) \in \mathcal Z \times \mathcal A, \mu \in S^{\sqcup}(\mathcal X)}$, and (iii) sampling the next queue state 
$(\xi_{i,x,\mu})_{x \in \mathcal Z, \mu \in S^{\sqcup}(\mathcal X)}$.

For the first step, fixing any $\pi$, we let
\begin{align*}
    \xi_{i,z} \sim \mathrm{Bernoulli}(\pi_{\lfloor t/3 \rfloor}(z)),
\end{align*}
for all $z \in \mathcal Z$, to sample action from $\pi$. 
Deferring for a moment the second step, for the new state in the third step, we use the Kolmogorov forward equation for the queue state during the epoch for any initial $(x, \mu)$: 
\begin{align*}
   P(x, \mu) = \exp{\mathbf Q(x,\mu)} \cdot e_{x_1} \in \Delta^{\mathcal Z} 
\end{align*}
with unit vector $e_{x_1}$,
which results from Poisson thinning, as the equivalent arrival rate at a queue of an agent currently in state $x$ with neighbourhood $\mu$ will be given by $\lambda \left(1 - x_2 + \frac{\mu(\mathbf 1_{\mathcal Z \times \{1\} \times S^{\sqcup}(\mathcal X)})}{d} \right)$. Thus, we have the transition rate matrix $\mathbf Q(x,\mu)$ with $\mathbf Q_{i,i-1} = \lambda \left(1 - x_2 + \frac{\mu(\mathbf 1_{\mathcal Z \times \{1\} \times S^{\sqcup}(\mathcal X)})}{d} \right)$, $\mathbf Q_{i-1,i} = \alpha$ for $i=1,\ldots,B$, and where $\mathbf Q_{i,i} = -\sum_j \mathbf Q(\lambda, z)_{j,i}$ for $i=0,\ldots,B$, and zero otherwise. Therefore, the next queue filling is sampled as
\begin{align*}
 \xi_{i,x,\mu} \sim \mathrm{Categorical}(P(x, \mu)).
\end{align*}
Lastly, for the second step we consider the non-random conditional expectation of packet losses during any epoch:
\begin{align*}
  \xi_{i,x_1,x_2,\mu} = [ \exp{\bar{\mathbf Q}(x,\mu)} \cdot e_{x_1} ]_{B+1}
\end{align*}
under the executed actions (allocations), analogously to the new state, by adding another row to the transition matrix $\mathbf Q(x,\mu)$, giving $\bar{\mathbf Q}(x, \mu)$ where $\bar{\mathbf Q}_{B+1,B} = \lambda \left(1 - x_2 + \frac{\mu(\mathbf 1_{\mathcal Z \times \{1\}})}{d} \right)$, i.e. counting all the expected packet arrivals whenever the queue is already full ($B$).

We use the same definitions for finite systems $(G_n, \mathbf z_n)$. We can verify \cite[Assumption A]{lacker2019local}, since $F$ is continuous e.g. by discreteness of the relevant spaces, and all $\xi_i$ are sampled independently and identically over agents and times. By \cite[Theorem 3.6]{lacker2019local}, the empirical distribution $\mu^{G_n, \mathbf z_n} \coloneqq \frac{1}{|G_n|} \sum_{i \in G_n} \delta_{X^{G_n, \mathbf z_n}_i(t)}$ converges in probability in $\mathcal P(\mathcal X^\infty)$ to its mean field limit $\mathcal L(X^{G,x}_{\varnothing})$, and in particular
\begin{align}
    \mu^{G_n, \mathbf z_n}(t) \to \mathcal L(X^{G,x}_{\varnothing}(t)) \quad \text{in distribution in $\mathcal P(\mathcal X)$}
\end{align}
at any time $t$, where $\mathcal P(\cdot)$ denotes the space of probability measures equipped with the topology of weak convergence. Hence, the above describes the original objective by
\begin{align*}
    &J^{G_n, \mathbf z_n}(\pi) 
     = \mathbb E \left[ \sum_{t=0}^\infty \gamma^t D^N(t) \right] \\
    &\quad = \sum_{t=0}^\infty \sqrt[3]{\gamma}^{t-2} \mathbb E \left[ r^\infty(\mu^{G_n, \mathbf z_n}(t)) \right] \\
    &\quad \to \sum_{t=0}^\infty \sqrt[3]{\gamma}^{t-2} \mathbb E \left[ r^\infty(\mathcal L(X^{G,x}_{\varnothing}(t))) \right] = J^{G, \mathbf z}(\pi)
\end{align*}
since we split any time step into three by the prequel, the continuous mapping theorem, and dominated convergence, where we use the continuous function 
\begin{align*}
    r^\infty(\mu) = \int \mathbf x_3 1_{x \in \mathcal Z \times \mathcal A \times [0, D_{\mathrm{max}}]} \mu(\mathrm dx),
\end{align*}
where $\mu = \mathcal P(\mathcal X)$,
to sum up the expected packet drops in the system, since the integrand is continuous and bounded (under the sum topology for the union in $\mathcal X$, and the product topology for products in $\mathcal X$).
\end{proof}

\begin{proof}[Proof of Corollary 2.2]
    Define the non-zero (by assumption) optimality gap:
    \begin{align*}
        \Delta J \coloneqq J^{G, \mathbf z}(\pi_i) - \max_{j \neq i} J^{G, \mathbf z}(\pi_j) > 0.
    \end{align*}
    Then, by Theorem~\ref{thm:Jconv}, there exists $n$ such that: 
    \begin{align*}
        \max_i \left| J^{G_n, \mathbf z_n}(\pi_i) - J^{G, \mathbf z}(\pi_i) \right| \leq \frac{\Delta J}{2}.
    \end{align*}
    Therefore,
    \begin{multline*}
        J^{G_n, \mathbf z_n}(\pi_i) - \max_j J^{G_n, \mathbf z_n}(\pi_j)  \\
        \geq \min_{j \neq i} (J^{G_n, \mathbf z_n}(\pi_i) - J^{G, \mathbf z}(\pi_i))
        + \min_{j \neq i} (J^{G, \mathbf z}(\pi_i) - J^{G, \mathbf z}(\pi_j)) \\
        + \min_{j \neq i} (J^{G, \mathbf z}(\pi_j) - J^{G_n, \mathbf z_n}(\pi_j))
        \geq - \frac{\Delta J}{2} + \Delta J - \frac{\Delta J}{2} = 0
    \end{multline*}
    is the desired conclusion.
\end{proof}


\section{Algorithm}
In order to solve the POMDP, we apply the established proximal policy optimization (PPO) RL method \cite{schulman2017proximal, yu2021surprising} with and without recurrent policies, as commonly and successfully used in POMDP problems \cite{ni2021recurrent}. PPO is a policy gradient method with a clipping term in the loss function, such that the policy does not take gradient steps that are too large while learning \cite{schulman2017proximal, yu2021surprising}. For our experiments, we have worked with the stable and easy-to-use RLlib 1.10 implementation \cite{liang2018rllib} of PPO. The overall training code is given in Algorithm~\ref{alg} given in \cite{}, which also shows how to analogously apply a trained MFC policy to a finite system. We use diagonal Gaussian neural network policies with $\tanh$-activations and two hidden layers, parametrizing MFC MDP actions $\zeta_t$ by values in $[0, 1]$ for each entry $\zeta_t(a \mid z)$, normalized by dividing by the sums $\sum_a \zeta_t(a \mid z)$.

\begin{algorithm}
    \caption{Learning MFC policies in finite systems}
    \label{alg}
    \begin{algorithmic}[1]
        \STATE \textbf{Input}: Hyperparameters and system parameters from Table~\ref{table:parameters} and Table~\ref{table:hyperparameters}.
        \STATE Initialize finite system, i.e. rates $\lambda_0 \sim \mathrm{Unif}(\{\lambda_h, \lambda_l\})$ and queue states $\mathbf z^n = \mathbf 0 \in \mathcal Z^{G_n}$ on topology $G_n$.
        \STATE Initialize PPO policy $\pi^\theta$, critic $V^\psi$.
        \FOR {PPO iteration $n = 0, 1, \ldots$}
            \STATE Initialize batch buffer $B = \emptyset$
            \WHILE[\textit{Execute on system}]{$|B| < B_b$} 
                \FOR {$t = 0, 1, \ldots, T_\mathrm e$}
                    \STATE Observe empirical state distribution $\mu^{G_n, \mathbf z_n}(t)$.
                    \STATE Sample decision rule $\zeta_t \sim \pi^\theta(\mu^{G_n, \mathbf z_n}(t)))$.
                    \FOR {$i = 1, \ldots, N$}
                        \STATE Observe agent state $z_t^{i}$ 
                        \STATE Sample action $a_t^{i} \sim \zeta_t(z_t^{i})$ (\textit{allocation rule}).
                    \ENDFOR
                    \STATE Execute allocation rules $\{a_t^{i}\}_i$ for $\Delta t$ time units.
                    \STATE Resample arrival rates $\lambda_{t+1} \sim \mathbb P(\lambda_{t+1} \mid \lambda_t)$.
                    \STATE Count number of dropped packets per agent $D_t$.
                    \STATE Observe new distribution $\mu^{G_n, \mathbf z_n}(t+1)$.
                    \STATE Save (state-action-reward-state) transition $B = B \cup \{ (\mu^{G_n, \mathbf z_n}(t), \zeta_t, -D_t, \mu^{G_n, \mathbf z_n}(t+1)) \}$
                \ENDFOR
            \ENDWHILE
            \STATE Compute GAE advantages $\hat A$ on $B$.
            \FOR[\textit{Train policy using PPO}] {epoch $i = 1, \ldots, T_b$}
                \STATE Sample mini-batch $b$ with $|b| = B_m$ from $B$.
                \STATE Update policy $\theta$ using PPO loss gradient $\nabla_\theta L_\theta$ on $b$.
                \STATE Update critic $\psi$ using $L_2$ loss gradient $\nabla_\psi L_\psi$ on $b$.
            \ENDFOR
        \ENDFOR
        \RETURN Policy $\pi^\theta$.
    \end{algorithmic}
\end{algorithm}

The rest of the system parameters and the PPO hyperparameter are given here in Tables \ref{table:parameters} and \ref{table:hyperparameters}, respectively.

\begin{table}
\footnotesize
    \centering
    \caption{System parameters used in the experiments.}
    \label{table:parameters}
    \begin{tabular}{@{}ccc@{}}
    \toprule
    Symbol     & Name          & Value     \\ \midrule
    $\Delta t$          &    Synchronization delay [ms]   & $1-10$ \\
    $\alpha$          &    Service rate [1/ms]  & 1   \\
    $(\lambda_h, \lambda_l)$          &    Arrival rates [1/ms] & $(0.9, 0.6)$   \\
    $N$          &    Number of agents   & $4 - 6000$   \\
    $M$          &    Number of queues  & $4 - 6000$   \\
    $d$          &    Number of accessible queues   & $3-6$   \\
    $I$          &    Monte Carlo simulations   & $100$   \\
    $B$          &    Queue buffer size   & $5-20$   \\
    $\mathbf z^n$ & Queue (agent) starting state & $  \mathbf 0 \in \mathcal Z^{G_n}$  \\
    $\nu_0$          &    Queue starting state distribution   & $[1, 0, 0, \ldots]$   \\
    $D$          &    Drop penalty per job   & $1$   \\ 
    $T$          &    Training episode length   & $50$   \\
    $G_n, G$          &    Graph topologies   & Section \ref{sec:topologies}  \\ \bottomrule
    \end{tabular}
\end{table}


\begin{table}
\footnotesize
    \centering
    \caption{Hyperparameter configuration for PPO.}
    \label{table:hyperparameters}
    \begin{tabular}{@{}ccc@{}}
    \toprule
    Symbol     & Name          & Value     \\ \midrule
    $\gamma$ &   Discount factor &  $0.99$\\
    $\lambda_\mathrm{RL}$ &   GAE lambda &  $1$\\
    $\beta_c$ &   KL coefficient & $0.2$ \\
    $\beta_t$ & KL target & $0.01$ \\
    $\epsilon$ &  Clip parameter & $0.3$ \\
    $l_{r}$ &   Learning rate & $0.00005$ \\
    $B_{b}$ &  Training batch size &  $4000 - 24000$ \\
    $B_{m}$ & SGD Mini batch size &  $128-4000$ \\
    $I_{m}$ & Number of SGD iterations & $5-8$ \\
    $T_b$ &  Number of epochs & $50$ \\  \bottomrule
    \end{tabular}
\end{table}
\section{Additional Experiments}

\begin{figure}
\centering
\includegraphics[width=0.8\linewidth]{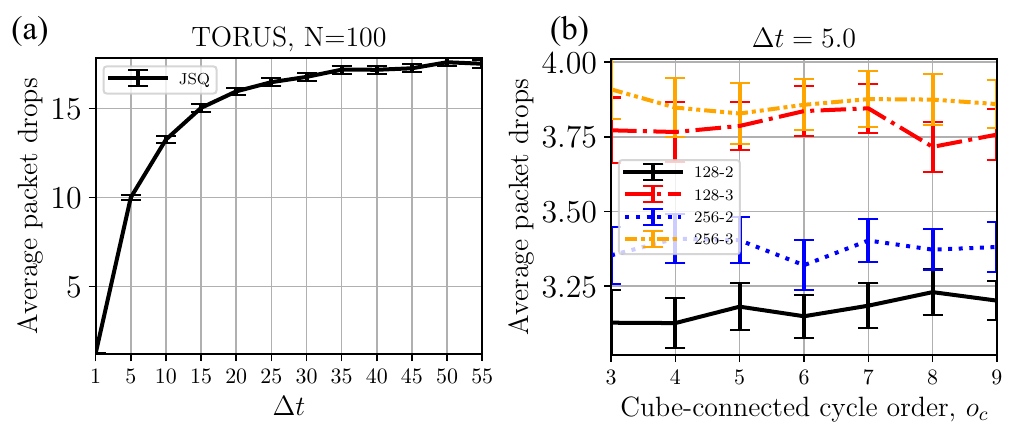}
\caption{(a) JSQ converges as $\Delta t$ increases, validating our simulator. (b) Performance evaluation on a fixed environment while changing only the neural network parameters. }
\label{fig:jsq_converge}
\end{figure}
\subsection{Simulator analysis} Firstly, we performed a small experiment in order to ensure that our simulator works as expected. Performance of the JSQ algorithm was evaluated for TORUS graph with $o_t=100$, $N=100$ and $d=4$ for the JSQ algorithm. It can be seen in Fig. \ref{fig:jsq_converge}(a) that on increasing $\Delta t$ there comes a point when the performance of JSQ starts to converge, which is the expected behaviour for finite systems. 

We also tried different sizes of neural network while keeping all other parameters and environment. The training was performed for the CCC graph with $o_c=5$ and $\Delta t=5$. The tested network sizes are shown in the legend of Fig. \ref{fig:jsq_converge}(b), where the first value is the size of the layer $(128,256)$ and the second value is the number of layers $(2,3)$ used. The performance was quite similar for all of them, and we used the default network size $[256-2]$ for all our experiments.
Finding more parameters could further improve the performance and will be investigated in the future.

\begin{figure}
\centering
\includegraphics[width=0.65\linewidth]{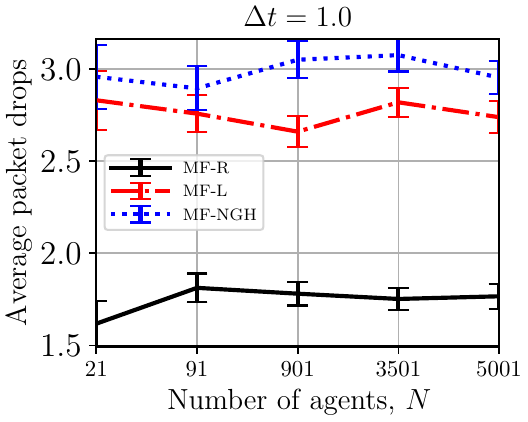}
\caption{MF-R is with symmetric actions, MF-L is learning actions for each neighbor separately but without knowing the exact state of the neighbor, MF-NGH is learning the action for each possible non-repeating combination of states for agents' own state and states of the neighbors. This experiment is done for the CYC-1D graph.}
\label{fig:diff-action-spaces}
\end{figure}

\subsection{Different policies} For the sake of completeness, we also compared different policies for the setup of CYC-1D graph at $\Delta t=1$. First is the MF-R where the policy learned tells you the probability with which to send to your queue or randomly to one of your neighbors. Second is the MF-L where the policy learned tells you the probability with which to send to your queue or ones of your neighbors. Lastly, is the MF-NGH where a separate action is learned for every possible combination of your state and state of your neighbors. However, the neighbors are still kept anonymized by not learning a separate policy for neighbors' state $[3,5]$ and $[5,3]$ for $d=2$, meaning that it does not matter which of your neighbors has queue filling $3$ or $5$, a single policy is learned for these combinations. In Fig. \ref{fig:diff-action-spaces} it can be seen that MF-R performs the best, and we believe this is because it has the smallest action space, so learning an optimal policy is more probable. Note that the input observation was the same for all, which is the empirical distribution of the agents' state as explained in Section 2.4.1. Hence, using any kind of action space is feasible, one just needs to consider the increase in training time and resources when the action space increases. Based on this we have used the MF-R model and policy for all our experiments.

\begin{figure}
\centering
\includegraphics[width=0.8\linewidth]{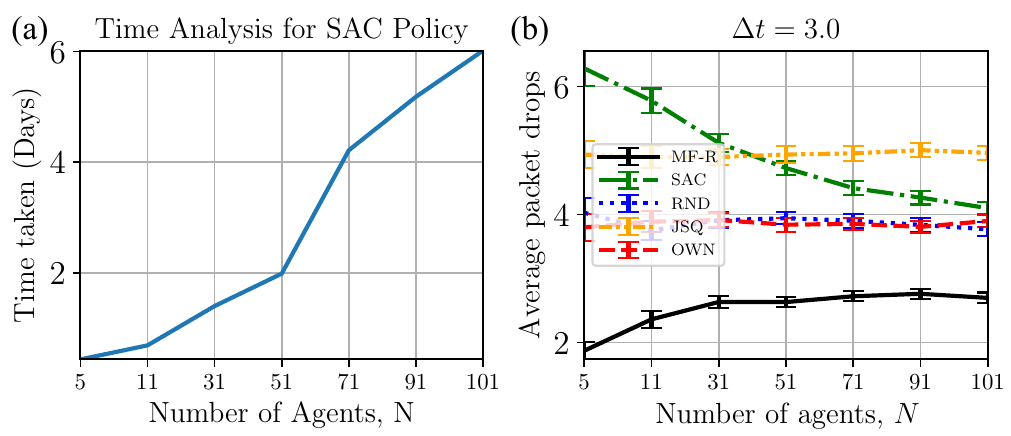}
\caption{(a): SAC training time to convergence increases with the number of agents, making it difficult to train and use for larger setups. (b): SAC performs worse than our proposed MF-R policy for a number of agents between $5$ and $101$.}
\label{fig:sac_comparison}
\end{figure}

\subsection{Comparison to SAC} The scalable-actor-critic (SAC) \cite{lin2021multi} method learns a discrete policy of sending to any one of the $d$ accessible queues. Each agent learns an individual policy while using as observation the agent's own queue state and also the queue states of all its neighbours. However, a trained policy cannot be used for an arbitrary number of agents since the policy of an agent is influenced by its neighbours states as well, which is not assumed to be the same for every agent. 
To begin, while training for SAC, we observed that on increasing the number of agents, the convergence to a locally optimal policy takes longer (using one core of Intel Xeon Gold 6130), making it not too feasible for the larger setups we consider in this work. See Figure~\ref{fig:sac_comparison}(a) for time taken to learn a SAC policy on a $1$-D cyclic topology with $\Delta t = 3$, $N=M$, and same computational resources for all training setups of $N$. Our implementation of SAC was adapted from \cite{yiheng2021}. Furthermore, Figure~\ref{fig:sac_comparison}(b) shows the performance of the learned SAC policy as compared to other algorithms. Although performance did improve as the number of agents rises, indicating the scalability of SAC to many agents, the performance of SAC remained suboptimal. Due to these limitations, we did not investigate the SAC algorithm further in our experiments.

\begin{figure}
\centering
\includegraphics[width=0.8\linewidth]{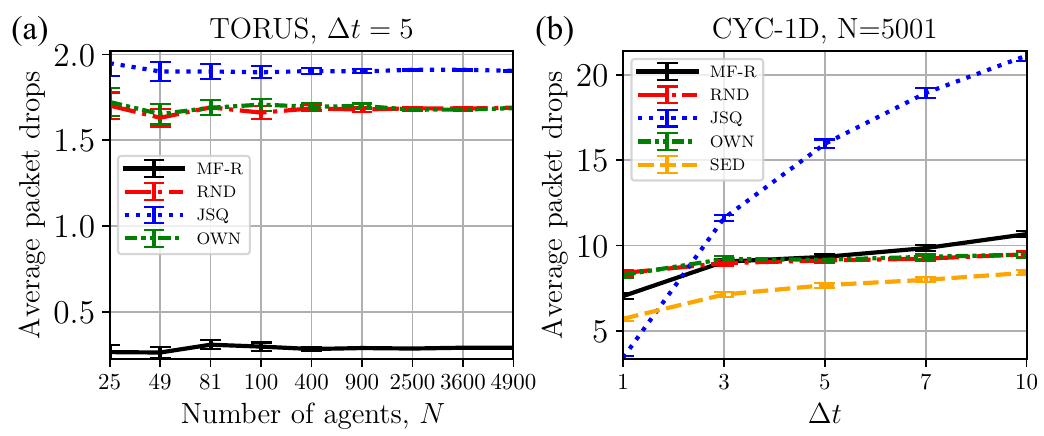}
\caption{(a) Performance comparison for increased buffer capacity of each queue to $20$. The system utilization was increased to ensure occurrence of packet drops in a limited amount of time. (b) Performance comparison of MF-R when the servers are considered to be heterogeneous. Additional comparison was done with the Shortest-expected-delay algorithm \cite{selen2016steady}, which is the state-of-the-art when the servers are heterogeneous. }
\label{fig:b20}
\end{figure}

\subsection{Increased buffer size} For completeness and to illustrate generality, we also performed an experiment in which the buffer size for each queue was increased from $5$ to $20$. To achieve faster convergence to a learned policy, we increased the arrival rate to the service rate of $1$. We also increased the time steps from $50$ to $200$ so that the queues can be sufficiently filled, and the policy can be learned over the increased state space. Fig. \ref{fig:b20}(a) shows that our learned policy outperforms the other algorithms at $\Delta t=5$ for TORUS graph.
However, this increased the training time.

\subsection{Heterogenous servers}
We also conducted experiments where we considered the servers to be heterogeneous with randomly assigned speed of  fast(rate $2$) or slow(rate $1$). The workload was the same; modulating between $[0.9, 0.6]$. Fig. \ref{fig:b20}(b) shows that our algorithm has comparable performance to RND and OWN while outperforming JSQ. We also compared with Shortest-expected-delay (SED), the state-of-the-art load balancing algorithm for heterogeneous servers \cite{selen2016steady}, which performs better since it makes decision based on both the neighbors' queue state information and the server speeds. We believe our model is not an ideal fit for this setting and further work needs to be done in this direction.

\begin{figure}
\centering
\includegraphics[width=0.8\linewidth]{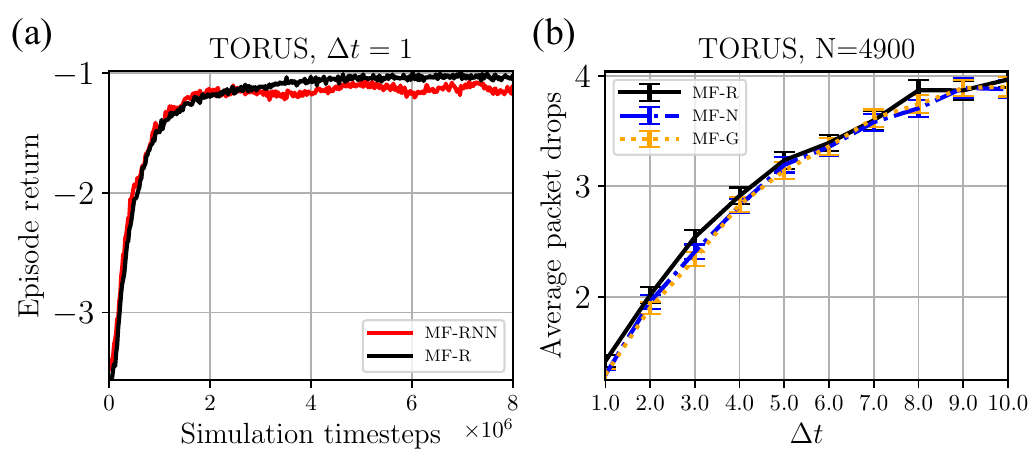}
\caption{(a): The training for TORUS with and without RNN policies converges almost to the same return. (b): Evaluation of the learned policy using different observations. MF-R uses only own queue state, MF-N additionally uses neighbour queue states, and MF-G uses state of all queues in the system.}
\label{fig:rnn}
\end{figure}

\subsection{RNNs and decentralized execution}
We additionally trained our MF-R policy with recurrent neural networks in PPO, as is typical for partially-observed problems \cite{ni2021recurrent}, MF-RNN. In Figure \ref{fig:rnn}(a), we see that their effect is negligible and can even be negative. Finally, we used partially observed decentralized alternatives to the empirical distribution as input to the learned MF-R policy in the evaluation, see Figure \ref{fig:rnn}(b). In MF-N we use the distribution of neighbours' queue states, in MF-G we use the empirical distribution of all queues in the system, and in MF-R only the agent's own queue state information is used as a one-hot vector. Similar performances indicate that the learned policy can be executed locally without using global information.


\end{document}